\documentclass{article}
\usepackage{multirow}
\usepackage{graphicx}
\usepackage{amssymb}
\usepackage{lineno}
\usepackage{amsopn}
\usepackage{colortbl}
\usepackage{array}
\usepackage[linesnumbered,ruled]{algorithm2e}
\usepackage{float} 
\DeclareMathOperator*{\argmax}{arg\,max}

\begin{document}
\title{Description and Retrieval of Geometric Patterns on Surface Meshes using an edge-based LBP approach}
\author{Elia Moscoso Thompson, Silvia Biasotti}
\maketitle
\begin{abstract}
While texture analysis is largely addressed for images, the comparison of the geometric reliefs on surfaces embedded in the 3D space is still an open challenge.
Starting from the Local Binary Pattern (LBP) description originally defined for images, we introduce the edge-Local Binary Pattern (edgeLBP) as a local description able to capture the evolution of repeated, geometric patterns on surface meshes. Our extension is independent of the surface representation, indeed the edgeLBP is able to deal with surface tessellations characterized by non-uniform vertex distributions and different types of faces, such as triangles, quadrangles and, in general, convex polygons. Besides the desirable robustness properties the edgeLBP exhibits over a number of examples, we show how this description performs well for 3D pattern retrieval and compare our performances with the participants to a recent 3D pattern retrieval and classification contest \cite{shrec17}.
\end{abstract}


\section{Introduction}
\label{sec:introduction}
3D content-based object retrieval and classification are receiving significant attention as the number and the size of 3D data is increasing. They potentially support the creation of search engines, 3D model catalogs, automatic 3D object segmentation, protein-docking, and so on.
Nowadays, a vast selection of methods exists, which is able to tackle 3D \emph{global} and \emph{local} similarity evaluations and \emph{coarse} and \emph{dense} shape correspondences \cite{Biasotti2016:CGF}.
Nevertheless, only few methods explicitly address the problem of retrieving and recognizing 3D patterns over surfaces \cite{othmani2013single,WerghiTBB15,WerghiBB15}. By 3D patterns, here we mean any geometric, repeated variation pattern over a surface embedded in the space. Far for being a purely academic question, classifying 3D patterns would have many practical applications: it would permit the automatic recognition of artwork patterns, including the attribution of the period of creation or an artist style \cite{Zeppelzauer:2016}; the classification of natural structures, such as like tree barks \cite{othmani2013single}; the identification of a surface or an object material, and so on. 
Recently, a benchmark for the automatic retrieval and classification of relief patterns over 3D scans of knitted fabrics has been launched at the SHape REtrieval Contest (SHREC) 2017 \cite{shrec17}. The performance of the methods till now run on that benchmark highlights that the retrieval and the classification of 3D relief patterns is still an open challenge.

A peculiar characteristic of 3D patterns is that they do not depend on the overall structure of the shape; rather, they identify parts or local properties that are independent of the global shape. 
Therefore, the recognition of 3D patterns must involve a good characterization of the local shape properties, which has to be: \textit{robust} to different model representations; \textit{sensitive} to the \emph{local} geometric variations that characterize the surface; as much as possible \textit{independent} of the surface bending, while keeping a reasonable computational complexity.

As the main contribution of this paper, we propose a novel extension to surfaces of the well-known Local Binary Pattern description \cite{ojala,ojala02} which is:
\begin{itemize}
\item able to deal with surface tessellations whose faces are made of convex polygons;
\item robust to non-uniform surface samplings;
\item invariant to object Euclidean transformations (roto-translations).
\end{itemize} 
With respect to a previous extension of the LBP to triangle meshes, the so-called meshLBP \cite{WerghiTBB15,WerghiBB15}, we base the LBP evaluation on the vertices of the tessellation, then we adopt a sphere-mesh intersection approach to determine the rings around a vertex and define a re-sampling criterion to obtain the same number of samples on each ring. 

Experimental results exhibit very good performances on various datasets and definitely overcome the algorithms proposed in the public contest on the retrieval of relief patterns at SHREC 2017 \cite{shrec17} showing the good potential of the proposed approach for real world applications.

The paper is organized as follows: Section \ref{sec:star} briefly overviews the recent literature on texture matching and 3D pattern retrieval; Section \ref{sec:background} introduces the basic concepts of our method, namely the LBP descriptor and the surface properties adopted to measure the geometric variations; Section \ref{sec:method} presents the \emph{edge Local Binary Pattern} (edgeLBP) method discussing the parameters involved in the description and its computational cost; Section \ref{sec:ex_settings} presents the datasets and the evaluation measures we used for our tests; Section \ref{sec:ex_results} discusses the robustness and the main properties of the edgeLBP description analyzing its performance over a set of surface meshes derived from laser scans of real objects and comparing, when possible, the edgeLBP with the meshLBP description; Section \ref{sec:shrec} reports the performances of the edgeLBP on the SHREC'17 pattern retrieval contest \cite{shrec17}; Section \ref{sec:realcosts} briefly describes the computational performances of the method. Discussions on the edgeLBP performance with respect to different types of patterns, open issues and future developments are provided in Section \ref{sec:conclusion}.
\section{Related work}
\label{sec:star}
Our problem can be seen as the natural extension in the 3D space of image texture description and retrieval. The literature on these topics is vast, therefore in this section we limit our references only to methods that are relevant to our approach, focusing on the main aspects of texture analysis and 3D pattern retrieval.

\paragraph*{Image texture retrieval}
Texture analysis has been largely addressed in computer vision and image processing and a variety of methods has been proposed, ranging from frequency-based to statistical-based methods \cite{review_texture}.
In general, the detection of patterns on real images is quite complex; the key aspect is the recognition of the texture properties robustly to the possible patterns variations \cite{cimpoi14}.

A typical strategy to detect patterns on images is to consider local patches that describe the behavior of the texture around a group of pixels. Examples of statistical descriptions are the Local Binary Patterns (LBP) \cite{ojala,ojala02}, the Scale Invariant Feature Transform (SIFT) \cite{Lowe2004} and the Histogram of Oriented Gradients (HOG) \cite{DaTr05}.

LBP-based methods are very popular and a large number of LBP variants has been proposed \cite{PietikainenHZA11}; for instance, addressing multi-resolution and  rotation invariance \cite{ojala02}, extending the definition to facial depth images \cite{Huang12}, human detection \cite{THANHNGUYEN2013} and also to volumetric images \cite{Fehr08,Citraro17}. An extended taxonomy  of 32 LBP variations and their performance evaluation for texture classification has been recently proposed in \cite{LIU2017} where the LBP is compared with 8 convolutional network based features over 13 datasets of 2D images. 
As an alternative to the volumetric LBP, the local frequency descriptor (LFD) \cite{Maani2014} is based on the gradient estimation on samples of a sphere around each pixel. Similarly to SIFT, only points that are detected as features are kept; then, the descriptors obtained in correspondence of these feature points are adopted to detect image anomalies such as brain anomalies and tumors in MRI images.

An aggregation of significant feature points obtained by pooling the point descriptors, e.g. SIFT+Fisher Vectors, can obtain significant texture classification performances \cite{cimpoi14}. Moreover, the combination of a SIFT-based feature description with Convolutional Neural Networks outperforms the feature-based descriptions on classic benchmarks approximately of the 10\% \cite{cimpoi16}.
Nevertheless, the problem of 2D texture recognition is still open because the learning strongly depends on the grouping of patterns which, in turn, is influenced by features that might strongly differ in type and size.

\paragraph*{Local feature descriptors on point clouds}
Methods like the Fast Point Feature Histograms (FPFH) \cite{Rusu:2009}, the SHOT descriptor \cite{TombariSS11}, the Spin Images \cite{Johnson:1999} and their recent extension named TOLDI \cite{TOLDI} mainly focus on point clouds. A quantitative analysis of the feature matching performance of local feature descriptors over standard datasets has been recently proposed in \cite{YANG2017}.
Unfortunately, most of these methods analyze the surface on the basis of its global appearance, discarding surface details and local shape variations.
For instance, the SHOT descriptor \cite{TombariSS11} is meant to solve point-to-point correspondences among sets of feature points. For this reason, these methods were adopted for facial matching and, in general, to address the partial similarity problem, focusing on the detection of feature correspondences rather than the comparison of surface patterns.

\paragraph*{3D pattern retrieval}
Methods in the literature for shape matching and retrieval 
can be classified according to their type of input, their local or global nature, their robustness to noise and model representations, their invariance to shape transformations and their suitability to partial matching \cite{veltkamp08,tam13,Biasotti2016:CGF}. For a detailed overview of algorithms and methods, we refer to recent surveys \cite{KaickZHC11,tam13,Biasotti2016:CGF} and to the proceedings of the annual SHape REtrieval Contest\footnote{http://www.shrec.net/} event. 
Here, we focus on methods that are potentially able to address the retrieval of 3D patterns, i.e., methods that are defined for surfaces, adopt a local shape description, are able to detected repeated features, and are independent of rigid transformations of the 3D models.

Partial similarity and, in particular, self-similarity are the key concepts currently referred to detect repeated, local features over a surface \cite{Liu13,YANG2017}.
For instance, the method \cite{gal2006salient} uses surface curvatures for recognizing salient shape features. Once these features are computed, they are mapped using a geometric hashing mechanism that determines the best transformation among these regions by mean of a voting scheme. The use of curvatures promotes the identification of well-detailed and isolated features encouraging the detection of shape details and (almost) flat regions while discards the whole shape structure. Such a technique is able to recognize repeated surface features (circles or stars) over a surface but, being based on geometry hashing, it is scale dependent and suffers of the local definition of “curvature” that could become insufficient when dealing with highly eroded or perturbed surfaces.
Similarly, \cite{Itskovich2011} observed that though isolated feature points often do not suffice, their aggregation provides adequate information regarding similarity. Then, the combination of segmentation techniques with the neighbor description of the feature points yields the detection of similar parts in bas-reliefs and archaeological artifacts. However, every surface part was considered as stand alone and no particular attention was allocated to the detection of repeated patterns. Moving further in this direction, the method proposed in \cite{Torrente2018} adopts the Hough transform to fit aggregated sets of feature points into template curves: while this approach naturally overcomes the problem of finding multiple instances of the same curve, it requires the surface can be locally projected on a plane and the vocabulary of possible curves is limited to those that have an algebraic expression.

When dealing with pattern characterization over surfaces (embedded in the Euclidean space), two strategies are possible: (i) to reduce the data dimension, i.e., to project the 3D data into an opportune plane (image) and apply an image pattern recognition algorithm to the projected data; (ii) to define the pattern description directly on the surface, fact which is not straightforward because it involves the treatment of three-dimensional data.

As a reference to the first typology of methods, we mention the method in \cite{othmani2013single} for tree species classification. There, the geometric variations of the tree trunk models are represented with a 3D deviation map over a cylinder that is flattened on a plane using the Principal Component Analysis (PCA) technique. Then, the geometric textures are compared using variations of the complex wavelet transform \cite{Kokare:2006}; see \cite{othmani2013single} for a detailed implementation analysis. Similarly, \cite{Zeppelzauer:2016} adopts a height map to project the reliefs and engraves of rock artifacts into an image and classify them. As further examples, we refer two approaches analyzed in the SHREC'17 contest \cite{shrec17} and labeled LBPI and CMC, respectively: namely, the LBPI \cite{shrec17} uses an image pattern method over a depth-buffer projection of the surface and the CMC \cite{shrec17} compares the principal curvatures in the mesh vertices using morphological image analysis techniques.

There are several generalizations to triangle meshes of image local feature description methods, such as the meshSIFT descriptor \cite{SMEETS2013}, the meshHOG descriptor \cite{meshHOG} and the textured Spin-Images \cite{Pasqualotto2013}.
The Mesh Local Binary Pattern (meshLBP) approach \cite{WerghiTBB16,WerghiTBB15,WerghiBB15} extends the LBP \cite{ojala} to triangle meshes and it is, at the best of our knowledge, the unique approach that  addresses 3D pattern classification and retrieval directly on the surface mesh. The main idea behind the meshLBP is that triangles play the role of pixels; there, the 8-neighbor connectivity of images is ideally substituted by a 6-neighbor connectivity around triangles. Rings on the mesh are computed by a uniform, triangle-based expansion. Working on (non-textured) meshes, the role of the gray-scale color is replaced by a function that is meant to capture the main pattern characteristics (in the examples, mainly Gaussian and mean curvatures, and the shape index \cite{Koenderink1992}). 
From the practical point of view, the meshLBP provides an efficient coding of a 3D pattern, providing a compact representation of the pattern. However, most properties on meshes are generally better represented on vertices instead of triangles (for instance the curvature value on a triangle is zero and usually it is approximated on vertices); moreover, triangles meshes can be really irregular: either in terms of connectivity or non-uniform vertex distribution. These facts jeopardize the efficacy of an expansion strategy based exclusively on elements of the mesh for detecting and coding representation-independent features.
\section{Basic concepts}
\label{sec:background}
We firstly summarize the \emph{Local Binary Pattern} (LBP) definition for images that we want to extend to surface tessellations. The salient point of the LBP is that it effectively codes the variations of the gray-level values, which are interpreted as a function (labeled $h$) over the image, around a pixel. Since we are interested to detect geometric variations we consider a set of curvature-based functions, detailed in Section \ref{sec:properties}.

\subsection{The Local Binary Pattern}
\label{sec:LBP}
The LBP is a reference description for texture recognition in still images \cite{ojala}. Among the LBP variations, here we introduce the terminology and the concepts we adopt in our paper.

\begin{figure}[t]
\centering
\scriptsize
\begin{tabular}{ccc}
\includegraphics[scale=.45]{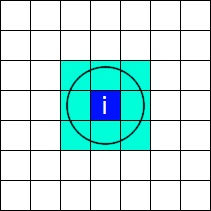} 	& 
\includegraphics[scale=.45]{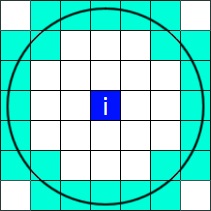}	&
\includegraphics[scale=.45]{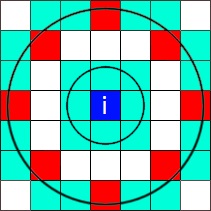}	\\
(a) $P=8$, $R=1.2$ & (b) $P=24$, $R=3.2$ & (c) $P=8$, $R=3.2$\\
\end{tabular}\\
\caption{(a-b): Rings with different radii $R$ relative to the pixel $i$. (c): Uniform down-sampling (from 24 to 8) for the pixels of the ring in (b). The down-sampled ring is represented by the red pixels.}
\label{fig:s_multiring}
\end{figure}

Let $I$ be a gray-scale image characterized by a pattern, $i\in I$ a pixel and $h$ the function such that $h(i)$ is the gray-level value of $i$. 
We denote $ring_1^i=\{i_1,...,i_8\}$ the set of 8 pixels adjacent to $i$, see Figure \ref{fig:s_multiring}(a). Usually, the ring $ring_1$ is clockwise ordered moving from the top left pixel. 

A binary string $str$ of 8 bits is associated to $i$ to code the variations of the gray-scale values of $i$ and the pixels in $ring_1^i$. For each $i_j \in ring_1^i$, the value $str(j)$ is defined as follows:
\begin{equation}
	str(j)=\Bigl\{
    \begin{array}{ll}
		1 & if \quad h(i)<h(i_j)\\
		0 & otherwise\\
    \end{array}
\end{equation}
The operator LBP labels the pixel $i$ with a scalar value derived from $str$ as follows:
\begin{equation}
	LBP(i)=\sum\limits_{j=1}^8 str(j)\alpha_k(j),
\end{equation}
where $\alpha_k$ is a weight function that determines the size of the descriptor. The most popular choices for $\alpha_k$ are: $\alpha_1(j)=1$  $\forall j$ and $\alpha_2(j)=2^j$  $\forall j$.

To achieve a multi-resolution description, it is possible to extend the LBP operator through a multi-ring coding \cite{ojala02}.
For each pixel $i$, a sets of concentric rings centered in $i$ ($ring_2^i$, $ring_3^i$, ..., $ring_{N_r}^i$) with increasing radii values is considered, see Figure \ref{fig:s_multiring}(a-b). Note that the rings are non necessarily square rings of pixels. Each ring is sampled with a pre-defined number $P$ of pixels, so that the string $str$ has the length $P$ on every ring, as shown in Figure \ref{fig:s_multiring}(c) for $P=8$. Finally, the multi-ring LBP descriptor is a matrix whose $k$-row corresponds to the LBP descriptor relative to $ring_k$.
\subsection{Curvature-based shape properties}
\label{sec:properties}
\begin{figure}[t]
\centering
\includegraphics[width=9cm]{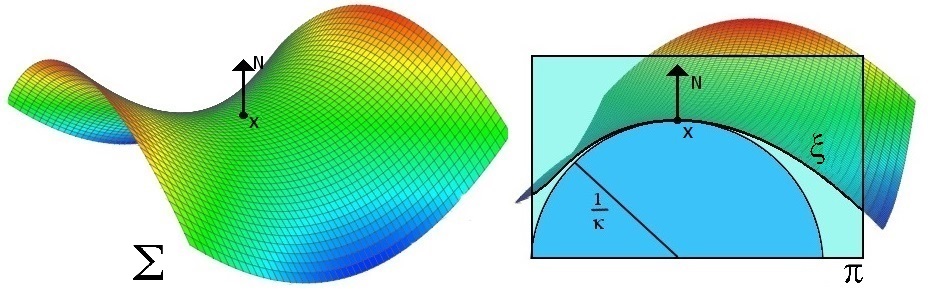}
\caption{Representation of the normal curvature to the surface $\Sigma$ in a point $x$. The oriented vector $N$ represents the normal vector in $x$. On the right, the surface is sectioned along the plane $\pi$ that generates the curve $\xi$. The radius of the osculating circle in $x$ relative to $\pi$ determines the normal curvature.}
\label{fig:normalcurvature}
\end{figure}
\begin{figure}[t]
\centering
\includegraphics[width=12cm]{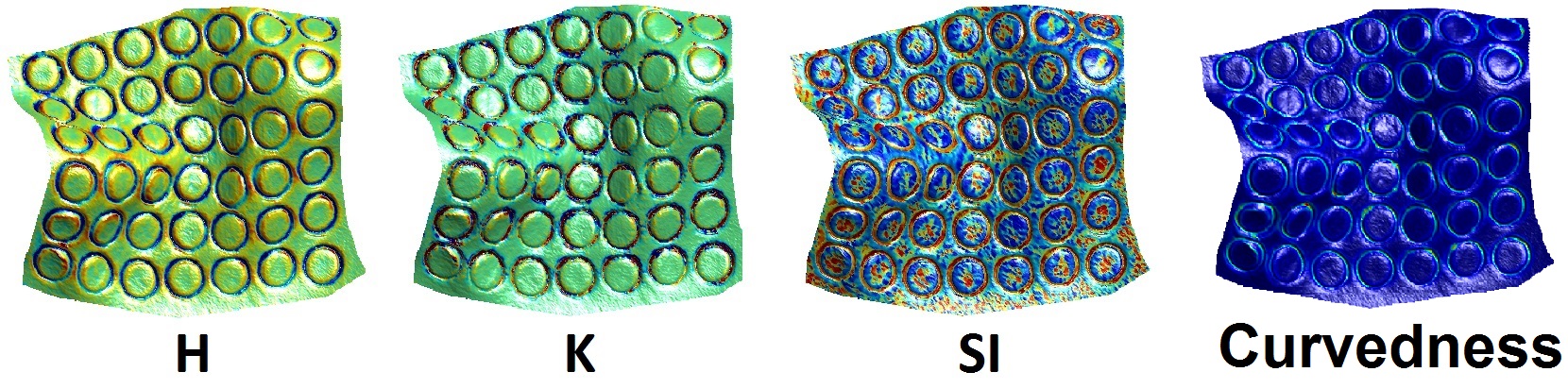}
\caption{Color representation of (from left to right) mean curvature, Gaussian curvature and SI. The lowest values are in blue, while greater ones fade from blue to green, then yellow and finally red.}
\label{fig:cur_functions}
\end{figure}

Geometrically, the curvature of a curve in a regular point is defined as the inverse of the radius of the osculating circle passing through this point, where, informally, by the osculating circle we mean the circle that fits the curve on a region infinitesimally small around that point. The extension of the concept of curvature to a surface $\Sigma$ is introduced through the notion of normal plane in a point. The intersection between the normal plane and $\Sigma$ defines an intersection curve $\xi$ where the curvature $\kappa$ in the point $x$ is well defined (and it is called normal curvature), see details in Figure \ref{fig:normalcurvature}. There are infinite intersection curves in a point $x$; however, the normal curvature assumes a minimum and a maximum value, denoted $k_1$ and $k_2$, respectively. $k_1$ and $k_2$ are called \emph{principal curvatures} of the surface in a point ($k_2\geq k_1$).
On the basis of the principal curvatures, other quantities are widely used to describe the local geometric differential properties of surfaces; in our experiments we considered also the mean curvature $H$, the Gaussian curvature $K$ , the Shape Index $SI$ and the $Curvedness$ \cite{Koenderink1992}, that are defined as follows:
$$H=\frac{k_1+k_2}{2}, \qquad 
K=k_1 \cdot k_2,$$ 
$$SI = \frac{2}{\pi}\arctan\left(\frac{k_1+k_2}{k_1-k_2}\right),\qquad 
Curvedness= \sqrt{\frac{k_1^2+k_2^2}{2}}.$$
Shape index is scale invariant and represents the local structure of a surface while the curvedness discriminates on the basis of the magnitude of the principal curvatures and, therefore, contains the scale-sensitive information. Studies in shape perception using smooth mathematical surfaces showed that the shape index and curvedness are measures that well reflect the human perception \cite{Phillips1996}.
A comparison of the algorithms for curvature estimation over surfaces having an analytical representation of the principal curvatures was proposed in \cite{vasa}. As confirmed by the experiments in \cite{vasa}, the method based on normal cycles proposed in \cite{Cohen03} provides a reasonable compromise between computational efficiency and quality of the curvature approximation and we opted for this approach. All the curvature quantities computed in this paper are approximated with the implementation of the curvature tensor proposed in the MATLAB toolbox \cite{ToolboxGraph}; following the default settings, we set to $3$, the size of the ring used to average the curvature tensor.  
Figure \ref{fig:cur_functions} depicts the values of $H$, $K$ and $SI$ approximated on a surface tessellation, using a color map that ranges from blue (low) to red (high) values.

For sake of completeness, we mention that, besides curvature measures, integral invariants are an alternate approach for the identification of convex, concave and flat regions. The volume integral originally proposed in \cite{Gelfand:2005} is related to the Gaussian curvature while the surface patch area behaves similarly to the mean curvature \cite{Mara:2017}. The advantage of these invariants is the adoption of numeric integration instead of simulating numeric differentiation; however, as discussed in \cite{Mara:2017}, they are very sensitive to surface details when dealing with the analysis of a tablet with cuneiform characters whose characteristics somehow resemble those of a 3D pattern.

\section{The edgeLBP description}
\label{sec:method}
\begin{figure}
\centering
\begin{tabular}{ccc}
\includegraphics[height=2.1cm]{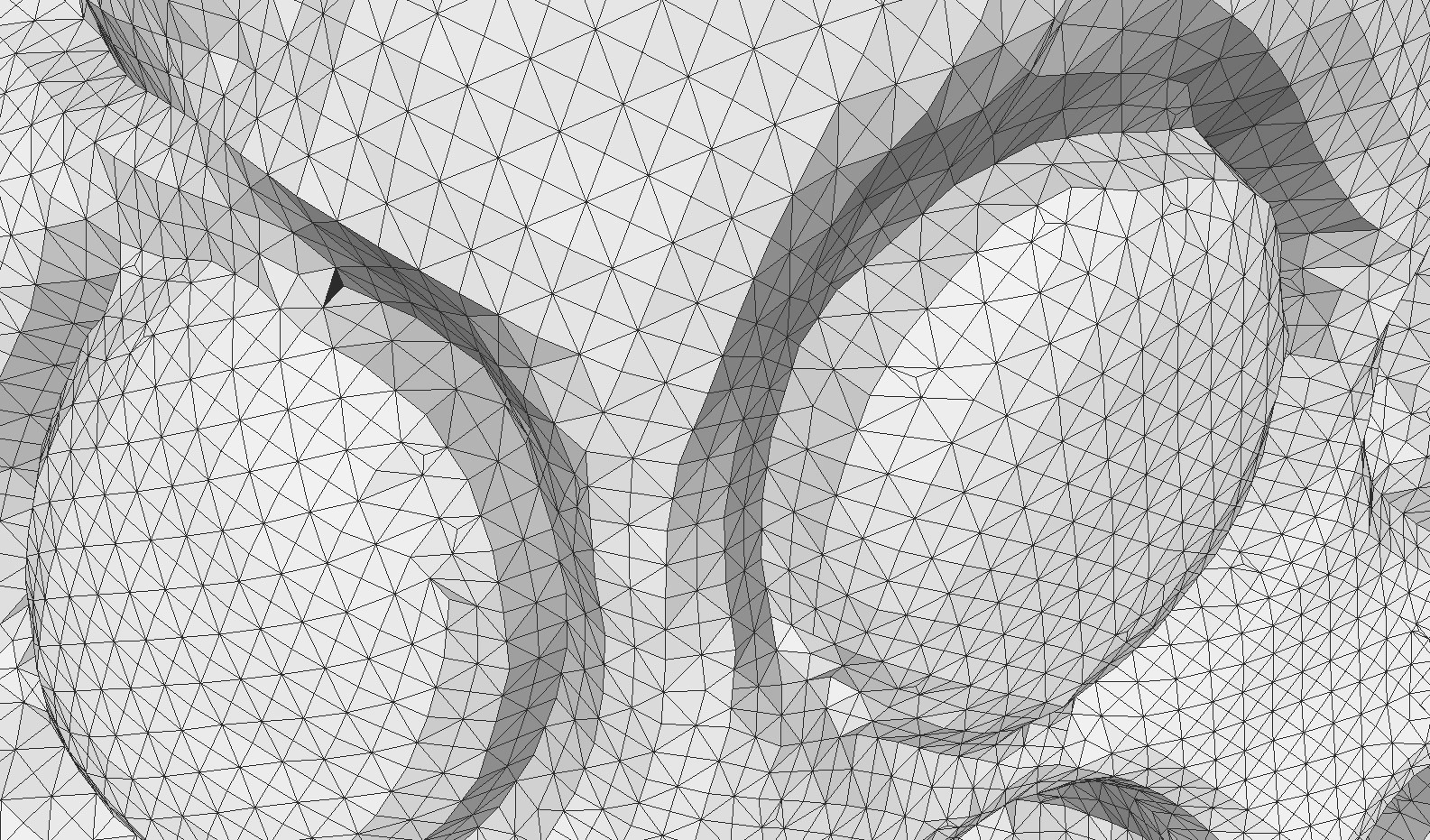} & 
\includegraphics[height=2.1cm]{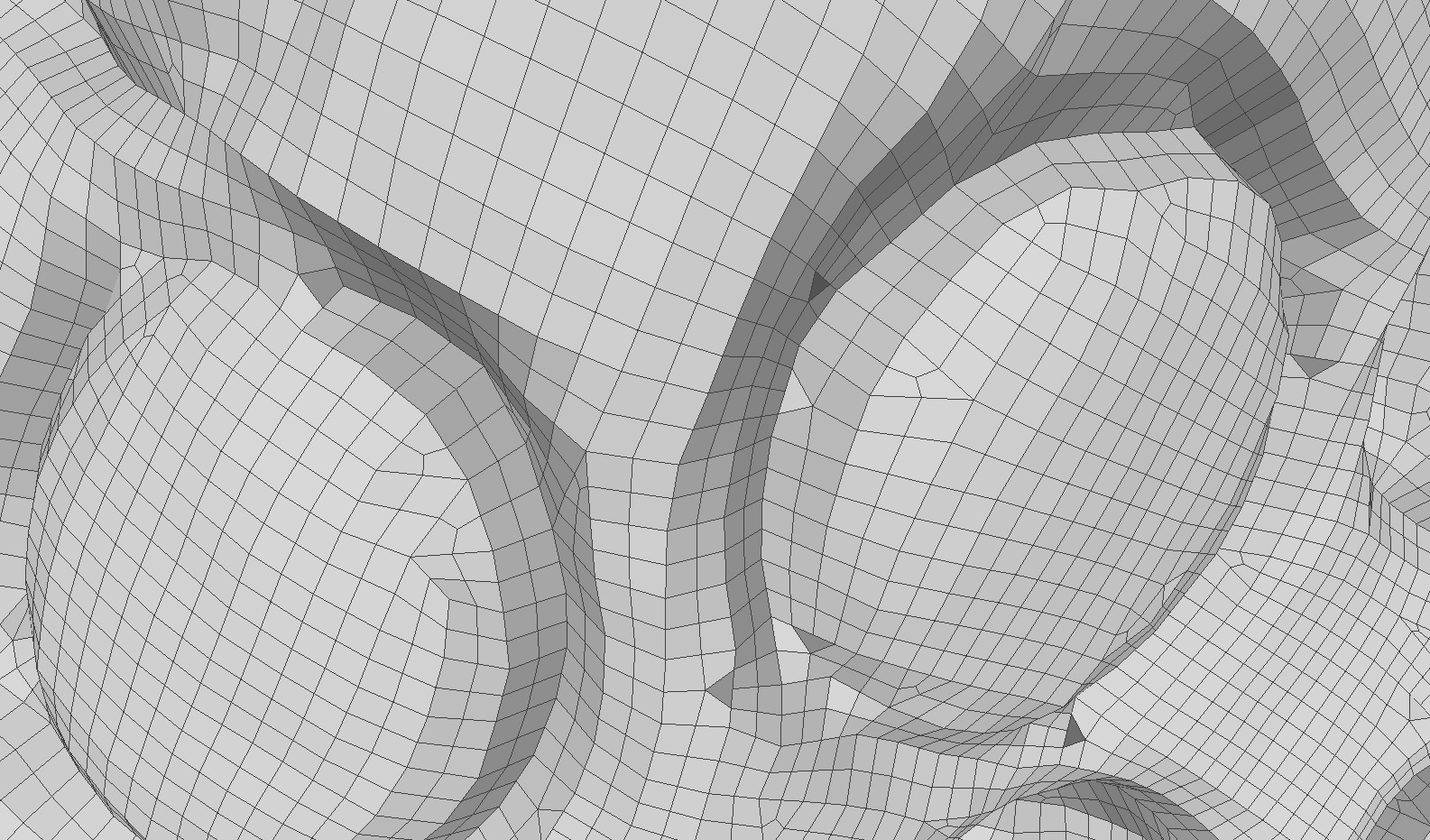} & 
\includegraphics[height=2.1cm]{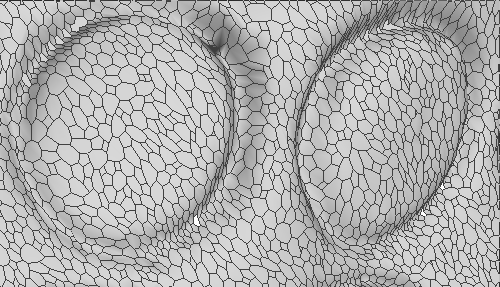}\\
(a) & (b) & (c)\\
\end{tabular}
\caption{Zoom on three surface tessellations: (a) a triangle mesh, (b) a quadrangle mesh, (c) a convex polygon mesh.}
\label{fig:meshes}
\end{figure}

The multi-ring LBP operator is extended to deal with surface tessellations using a sphere-mesh intersection technique, see Section \ref{sec:edgeLBP}. 
With the term \emph{surface tessellation}, we mean a polygon mesh $T=(V,E,F)$ which is a collection of vertices $V$, edges $E$ and faces $F$ that defines the surface of a polyhedral object. In our settings, we also assume that the faces are convex polygons. Popular examples of surface tessellations are triangle meshes, quadrangulations, and the Centroidal Voronoi Tessellations (CVT) \cite{Du:1999}, some examples of possible surface representation are shown in Figure \ref{fig:meshes}.
We also assume that each pattern property can be coded as a scalar function $h$ defined on the vertices of the tessellations, formally, $h:V\rightarrow\mathbb{R}$.

Our algorithm is detailed  for generic tessellations in the Sections \ref{sec:edgeLBP} and \ref{sec:rings}, while Section \ref{sec:parameters} details how to set the edgeLBP parameters. 
Section \ref{sec:distances} describes how to pass from a local description to a global pattern description and how to use the edgeLBP description to define the dissimilarity measure between two tessellations. Finally, Section \ref{sec:costs} exhibits the computational complexity of the edgeLBP technique.

\subsection{Multi-Ring Sampling}
\label{sec:edgeLBP} 
While a pixel grid has the same connectivity anywhere, surface tessellations can be widely \emph{irregular}. By irregular we mean that the vertices can be non uniformly distributed over the surface. Furthermore, the faces of the tessellation may have different area, shape and number of edges.

\begin{figure}[t]
\centering
\begin{tabular}{cccc}
\includegraphics[height=3cm]{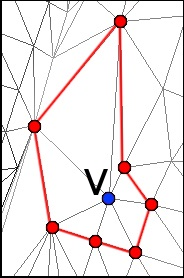} &
\includegraphics[height=3cm]{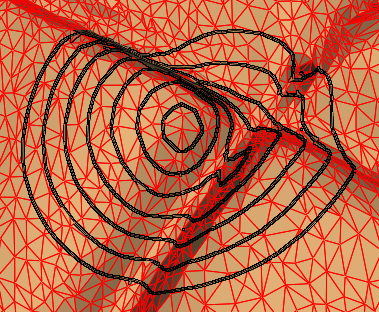} &
\includegraphics[height=3cm]{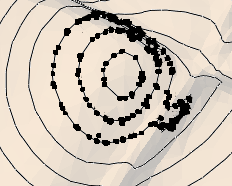} & 
\includegraphics[height=3cm]{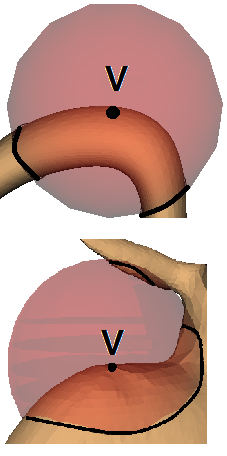}\\
(a) & (b) & (c) & (d)\\
\end{tabular}
\caption{(a): The ring of the vertex $v$ (in blue) formed by vertices over a triangle mesh; (b): multiple rings of a single vertex are shown; the black dots in (c) ($p_i$ in our notation) represent the intersections between black rings in (b) and the edges of the mesh; (d), from top to bottom: first, $S^v_{k}$ corresponds to the connected component that contains $v$, the other component is discarded; second, $S^v_{k}$ is non-simply connected and $v$ is non-admissible.}
\label{fig:ring_nature}
\end{figure}

As discussed in Section \ref{sec:LBP} the notion of ring is crucial for the LBP description. In case of triangle meshes, an intuitive transposition of the notion of ring would be a ring defined as a set of vertices. In Figure~\ref{fig:ring_nature}(a) we show the ring of the vertex $v$ formed by the sequence of red edges and vertices. Indeed, the irregularity of the mesh elements strongly influences rings defined on mesh elements only, because it would not be invariant to different tessellations of the surface, even simple edge swaps. Rings that are associated to different elements of the tessellation could carry information about surface portions with significantly different shape.

To overcome these limitations, given a surface tessellation $T$, we define the ring of a vertex $v \in V$ as the intersection of the surface tessellation with a sphere centered in $v$ of a given radius $R$. Then, we look at the intersections $p_i$ between the sphere and the \emph{edges} of the tessellation, creating a set of points  $\mathcal{R}=\{p_1, p_2, \ldots, p_k\}$, that approximates the curve which is the intersection between the sphere and the surface. We linearly interpolate the set $\mathcal{R}$ to obtain a continue and closed curve $C$ that represents the ring; $C$ is oriented counter-clockwise with respect to the vector in $v$ normal to $T$.
The value $h(p_i)$ on the points of $\mathcal{R}$ is determined by the weighted average of the value the function $h$ assumes on the vertices that limit the edge $e \in E$ such that $p_i \in e$.
Generally the number of elements of $\mathcal{R}$ varies from one ring to another, because of the increasing radius and the irregularity of the tessellation. To keep the number of elements constant on every ring, we sample $C$ with a fixed number of points $P$; we call $P$ the \emph{spatial resolution}.

To achieve a multi-ring representation, for any vertex $v \in V$ we consider $N_r$ rings, $\{ring_1^v, \ldots, ring_{N_r}^v\}$. Let $S^v_{k}$ be the surface portion of $T$ that contains $v$ and has the $ring^v_k$ as its boundary, $k=1\ldots N_r-1$, then the relation $S^v_{k} \subset S^v_{k+1}$ holds for each $k$. 
We take advantage of this relation to optimize the sphere-tessellation intersection adopting a region growing expansion around the vertex $v$. Examples of the intersection of the sphere at increasing radius around a vertex are shown in Figure \ref{fig:ring_nature}(b-c), while details on the algorithm are provided in Section \ref{sec:rings}. 

Similarly to the standard LBP approach and to avoid a possible ambiguity close to the surface boundaries, we consider as \emph{admissible} only the vertices for which all the $N_r$ rings are closed curves $C$.

In general, the sphere-surface intersection can produce multiple, closed curves that bound either a multiple connected or a dis-connected portion of surface; some examples are shown in Figure \ref{fig:ring_nature}(d), for a detailed vertex classification based on a sphere-mesh intersection approach we refer to \cite{Mortara2004}.
Using a region growing approach, $S^v_{k}$ is the portion of the sphere-surface intersection that contains $v$. 
If the boundary of $S^v_{k}$ is a closed curve, $v$ is considered an admissible vertex, otherwise it is \emph{non-admissible} for the edge-LBP.
Note that with the edge-LBP we are interested to code local geometric variations on the surface (like corrugations, incisions, and so on), therefore the radius $R$ should be kept small with respect to the overall dimension of the surface.
This implies that the choice of the radius $R$ is crucial for the type (and the size) of the patterns we are going to identify; indeed it must be not too large to avoid to mix global and local surface information and not too small to be significant.
In practice, multiply-connected regions appear only in case of topological noise, like small handles and self-intersections of the mesh and in our experiments over thousands of tessellations we never met admissibility problems.

\subsection{Implementation}
\label{sec:rings}
What follows details the routines adopted to evaluate the edgeLBP over a single ring 
of a vertex, then we outline how to extend it to a multi-ring representation. Overall, we identify three main steps:
\begin{enumerate}
\item \emph{RingExtraction} - The ring $\mathcal{R}$ given by the intersection between the sphere of radius $R$ centered in $v$ and the tessellation edges is computed according to the procedure detailed in Algorithm \ref{pcode:ring_pc}. The function \emph{VE(v)} calls the basic function to the data structure that returns the list of all the edges that are incident to the vertex $v$. Once the edges that intersect the sphere with center $v$ of radius $R$ are identified (lines 2-15 of Algorithm \ref{pcode:ring_pc}), the coordinates of the intersection points $p_i$ are computed (the function \emph{EdgeSphereIntersection} numerically evaluates the intersection of an edge with a sphere). 

We propose a method to sort $\mathcal{R}$ with respect to a starting point $\tilde{p}$. $\tilde{p}$ is selected according to a shape-based criterion and therefore is rotation and translation invariant. Starting from $\tilde{p}$ the function \emph{SortingP} counter clock-wisely sorts the points of $\mathcal{R}$ with respect to the normal to the surface in $v$. Even if this sorting would not influence the $\alpha_1$ representation we adopt in the paper, it would become crucial if considering the $\alpha_2$ one.
The point $\tilde{p}$ verifies the relation: 
$$\tilde{p}=\argmax\limits_{p_i\in\mathcal{R}} h(p_i).$$
In  Figure \ref{fig:start_point_test} we detail the choice of $\tilde{p}$ for three meshes that correspond to three different resolutions of the same model. There, for each vertex we depict the unit vector defined as $\overrightarrow{n}=\frac{\overrightarrow{\tilde{p}-v}}{||\tilde{p}-v||}$.
\begin{figure}[ht]
\centering
\begin{tabular}{ccc}
$20k$ & $10k$ & $5k$ \\
\includegraphics[width=3.6cm]{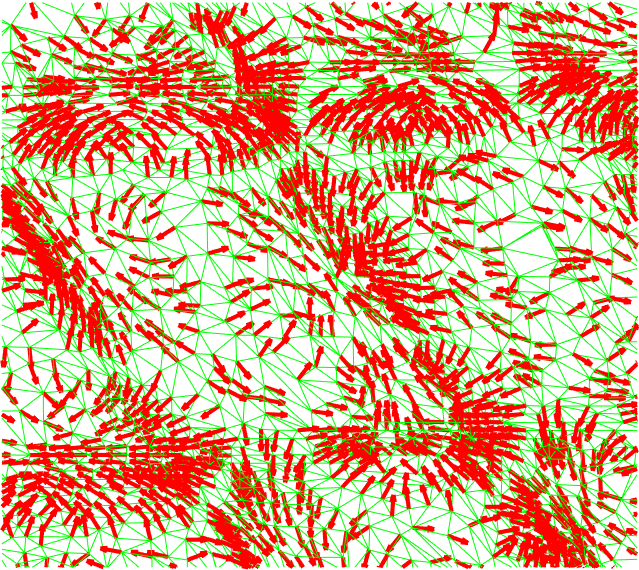}&
\includegraphics[width=3.6cm]{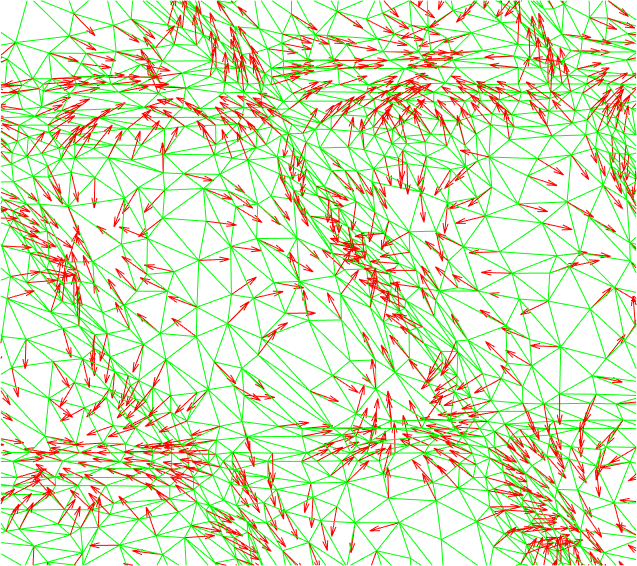}&
\includegraphics[width=3.6cm]{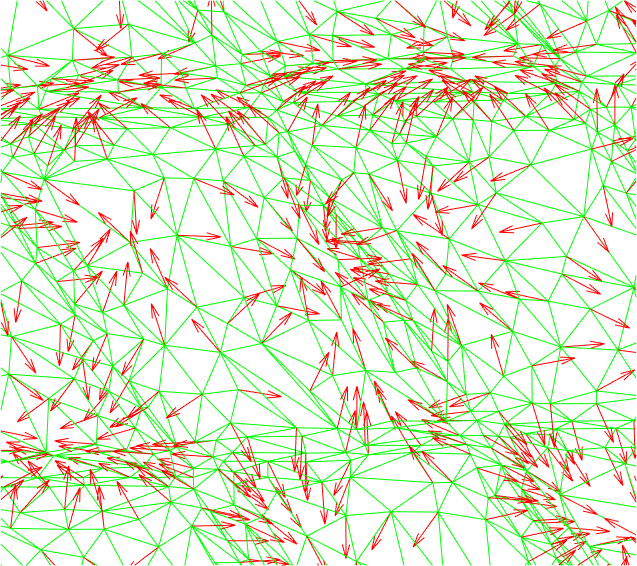}
\end{tabular}
\caption{Arrows represent the orientation of vector that connect a vertex $v$ with the starting point of the rings centered in $v$; from left to right, details on a surface mesh of $20K$ vertices and two re-samplings with $10K$ and $5K$ vertices: the choices of $\tilde{p}$ is robust to mesh decimation and depends on the local geometry of the surface.}
\label{fig:start_point_test}
\end{figure}
As expected, the starting point of the rings is stable in most of the vertexes of the mesh. This fact was confirmed in numerous experiments we performed on meshes of different resolution. In case of symmetries around a vertex, multiple choices of the starting point are possible: we select the candidate point that is the farthest from the other elements of $\mathcal{R}$.

\item \emph{RingResampling} - 
Given a ring $\mathcal{R}$ representing a simple, closed curve we identify $P$ equidistant samples $s_k$, $k=1, \ldots, P$ on $\mathcal{R}$ as detailed in Algorithm \ref{pcode:equis_pc}. The values $h(s_k)$ are linearly approximated (function \emph{Sample} in Algorithm \ref{pcode:equis_pc}) as follows: denoting $p_i$ and $p_{i+1}$ the two consecutive points of $\mathcal{R}$ on which the sample $s_k$ falls, the value $h(s_k)$ is equal to the weighted mean of $h(p_i)$ and $h(p_{i+1})$.
At the end of this procedure, the values $h(s_k), k=1, \ldots, P$ are returned in the array $S$.
\item \emph{edgeLBP Evaluation} - Once the value of the function $h$ is known on the sample set $S$, the evaluation of the edgeLBP on the vertex $v$ is straightforward. Here the function $\alpha$ represents one of the weight functions $\alpha_1$ or $\alpha_2$ defined in Section \ref{sec:LBP}, see lines 2-8 of Algorithm \ref{pcode:elbp_comp}.
\end{enumerate}
\begin{algorithm}[t]
\small
\DontPrintSemicolon
\SetKwInOut{Input}{Input}
\SetKwInOut{Output}{Output}
\SetKwInOut{return}{return}
\Input{A tessellation $T=(V,E,F)$, a vertex $v \in V$, a radius $R >0$.}
\Output{A set $\mathcal{R}$ of samples $p_i$ of the ring with center in $v$ of radius $R$.}
\Begin{
$L \leftarrow \emptyset$\;
$U \leftarrow VE(v)$\;
\While{$ U \neq \emptyset$}{
\For{$e=(v1,v2)\in U$}{
\If{$(d(v1,v)-R)*(d(v2,v)-R)<0$}{$L\leftarrow L\cup \{e\}$\;}
\If{$(d(v1,v)-R)<0\quad {\bf or}\quad (d(v2,v)-R)<0$}{$U\leftarrow U \cup VE(v_1) \cup \ VE(v_2)$}
mark $e$\;
}
$RemoveMarked(U)$
}
\For{$e \in L$}{
$\mathcal{R} \leftarrow EdgeSphereIntersection(e,R,v)$ \;
}
\Return{$\mathcal{R} \leftarrow SortingP(\mathcal{R},T)$}
}
\caption[caption]{RingExtraction.\\\hspace{\textwidth} \emph{Notes}: $d$ is the euclidean distance, $U$ is the list of the edges that may have a $p_i$ on them (initially empty), $L$ is the list of edges the algorithm has already checked.}
\label{pcode:ring_pc}
\end{algorithm}

When extending the edgeLBP evaluation to multiple rings, the \emph{RingExtraction} procedure is modified to take advantage of the nested nature of the rings; i.e., rings are computed increasingly with respect to the radius $R$. The initialization in Algorithm \ref{pcode:ring_pc} of the set $U$ of edges that are suitable for the sphere-surface intersection, starts from the edges that originated the previous ring and does not take into account edges already visited. 
Moreover, only the biggest ring $ring_{N_r}$ is sorted as described in Section \ref{sec:rings}: we sort all the other rings centered  in the vertex $v$ consistently this sorting. 
In particular, we consider the plane $\pi$ passing through $v$ with $\overline{w}=n(v)\times \overline{(v-\tilde{p})}$ as its directional vector, where $n(v)$ is the normal, unit vector to the surface in $v$ and $\tilde{p}$ is the starting point of $ring_{N_r}$. Then, we choose as the starting point on each ring the closest point to $\tilde{p}$ and order all the rings counterclockwise with respect to the normal in $v$. 

In our settings, we opted for a uniform distribution of the ring radii. For instance, denoting $R_{max}$ the maximum radius is will be $\frac{R_{max}}{N_r}, 2\frac{R_{max}}{N_r}, \ldots, R_{max}$.
\begin{algorithm}[t]
\small
\DontPrintSemicolon
\SetKwInOut{Input}{Input}
\SetKwInOut{Output}{Output}
\SetKwInOut{return}{return}
\Input{A set $\mathcal{R}$ of intersection points, the function $h$, the spatial resolution $P$.}
\Output{An array $S$ of $P$ scalar values $s_k$.}
\Begin{
$length\leftarrow \sum d(p_i,p_i+1)$\;
$dl=\frac{length}{m}$\;
$idx_{end}\leftarrow 2$\;
$index\leftarrow 1$\;
$s(1)\leftarrow h(p_1)$\;
\While{$size(S) \neq m$}{
\While{$d_{R}(p_1,p_{idx_{end}})-dl\cdot index \leq 0$}{$idx_{end}$++}
$index$++\;
$S\leftarrow Sample(h(p_{idx_{end}-1}),h(p_{idx_{end}}))$\;
$idx_{end}$++\;
}
\return{S}
}
\caption{RingResampling\label{pcode:equis_pc}}
\end{algorithm}
\begin{algorithm}[t]
\small
\DontPrintSemicolon
\SetKwInOut{Input}{Input}
\SetKwInOut{Output}{Output}
\SetKwInOut{return}{return}
\Input{The array $S$, the pivot value $h(v)$.}
\Output{The value edgeLBP($v$).}
\Begin{
\For{$idx=1:numel(S)$}{
\eIf{$h(S(idx))<h(v)$}{$str(idx) \leftarrow 0$\;}{$str(idx) \leftarrow 1$\;}

}
\return{$edgeLBP(v)\leftarrow \sum str(j)\alpha(j)$}
}
\caption{edgeLBP Evaluation\label{pcode:elbp_comp}}
\end{algorithm}
\subsection{Parameter settings}
\label{sec:parameters}
While the choice of the number of rings $N_r$ follows the classic LBP approaches, the values of $R_{max}$ and $P$ are set on the basis of the following reasonings:
\begin{itemize}
\item $P$ corresponds to the number of samples over each ring. In case of a circle on a flat surface, it would correspond to the number of sectors that would divide the angle $2\pi$. Based on our tests, this parameter should be included between 12 and 18 (for flat surfaces, this would correspond to a uniform sampling with an angle that ranges from $\frac{\pi}{10}$ to $\frac{\pi}{6}$);
\item $R_{max}$ represents the radius of the biggest sphere used to define the rings. It can be chosen by the user on the basis of the size of the variations (patterns) to be coded.
Nevertheless, we also suggest two possible automatic ways to define $R_{max}$, both based on the assumption that a pattern on a surface should be quite small with respect to the global size of the model. Namely:
\begin{itemize}
\item $R_{max}=\frac{1}{10}\sqrt{\frac{A}{\pi}}$. This is a scale-invariant radius based on a fraction of the area of the whole surface, where $A$ represents the area of the surface model. 
\item $R_{max}=C \cdot el$. This is a calibration of the radius based on the average length of the tessellation edges $el$ and $C$ is a constant, $C\in [10,20]$.
\end{itemize}
\end{itemize}
\subsection{edgeLBP description and similarity measure}
\label{sec:distances}
Given the surface tessellation $T$, its edgeLBP descriptor $D_T$ is defined as a feature vector; in particular, the value $DT(n,m)$ corresponds to the number of vertices that assume edgeLBP value $m$ on the $ring_n$. The size of $DT$ is equivalent to $N_r(P+1)$.
Since in the experiments we are mostly interested in a probability histogram of the distribution of the edgeLBP values, we adopt $\frac{D_T}{n_v}$ as the edgeLBP descriptor, where by $n_v$ we mean the cardinality of the set $V$ of the vertices of $T$. Through this normalization of $T$ we achieve robustness to the number of vertices of the surface representation. 

We define the (dis)similarity between two tessellations $A$ and $B$ as 
the distance between their corresponding edgeLBP descriptors $D_A$ and $D_B$. 
Since the edgeLBP can be thought as a matrix, any feature vector distance is suitable to evaluate the similarity between two edgeLBP descriptions. In the experiments shown in this paper, we adopt the Bhattacharyya  and the $\chi^2$ distances \cite{Deza2009}, which are widely used in image processing. 
For the discrete case, the Bhattacharyya distance between two distributions $\phi$ and $\psi$ of a scalar random variable $X$ has the following formulation:
$$d_{Bha}(\phi,\psi)=\sqrt{1-BC(\phi,\psi)}, \qquad BC(\phi,\psi)=\sum\limits_{x\in X} \sqrt{\phi(x)\psi(x)}, $$
where $BC$ is called the \emph{Bhattacharyya coefficient}.
We also tested other distances (like the Euclidean distance and the Earth Mover's Distance \cite{Deza2009}) but the results obtained with the Bhattacharyya and the $\chi^2$ distances provided the best performances. In most of the experiments, the Bhattacharyya and the $\chi^2$ distances behave equivalently, when different we specify in the text the distance adopted.

For a set of surface tessellations, the (dis)similarity values are stored in a \emph{distance matrix} $Dist$, where $Dist(i,j)=d_{Bha}(D_i,D_j)$ is the distance between the descriptor of the tessellation $i$ and $j$. The diagonal values of $Dist(i,i)$ are zero.
\subsection{Computational cost}
\label{sec:costs}
Given a surface tessellation $T$ with $n_v$ vertices, we briefly discuss the computational complexity of the routines involved in the edgeLBP evaluation.

We assume that the tessellation in input is stored in an appropriate data structure, therefore the cost of computing the relations among the elements of the tessellation (e.g., vertex-edge, face-vertex, vertex-face) is constant or $O(n_v)$, depending on the relations. Also, the shape properties are precomputed: the curvature
estimation proposed in \cite{Cohen03} has computational complexity $O(n_v \log n_v)$.

If the tessellation $T$ is with boundary, the creation of the list of the vertices that are admissible for the edgeLBP operator is based on the distances of the vertices from the boundary. This preprocessing phase costs $O(n_v\log n_v)$ operations.

For each vertex which is admissible, the computation  of the intersection between a sphere and the edges of the tessellation has complexity $O(n_v)$ (in the worst case), and, therefore, the cost is $O(n_v^2$) for the whole surface.
It's worth noticing that this cost holds if the sphere intersection processes all the vertices of the tessellation for every vertex. In average, the radius of the sphere is considerably small and diminishes the average computational complexity to $O(n_v \log n_v)$.
Ring re-sampling has linear complexity with respect to the number of elements of the rings; in the worst case the number of elements in the rings of a vertex $v$ can become $O(N_r \cdot n_v)$, where $N_r$ is the number of rings (and it is constant). Thus, the re-sampling of all the rings costs at most $O(n_v^2)$ operations and, in average, the computational complexity is again $O(n_v \log n_v)$.

Finally, the computation of edgeLBP histogram is linear in the number of samplings of the tessellations that are O($N_r \cdot P \cdot n_v)$; since $N_r$ and $P$ are two parameters that are constant, the cost is $O(n_v)$. The time performance of the algorithm for real data is provided in Section \ref{sec:realcosts}.
\section{Experimental environment}
\label{sec:ex_settings}
This Section lists the datasets and the measures used to evaluate the retrieval and classification performance of the edgeLBP description.
\subsection{Datasets}
\label{sec:database}
\begin{figure}[!t]
\centering
\includegraphics[width=12cm]{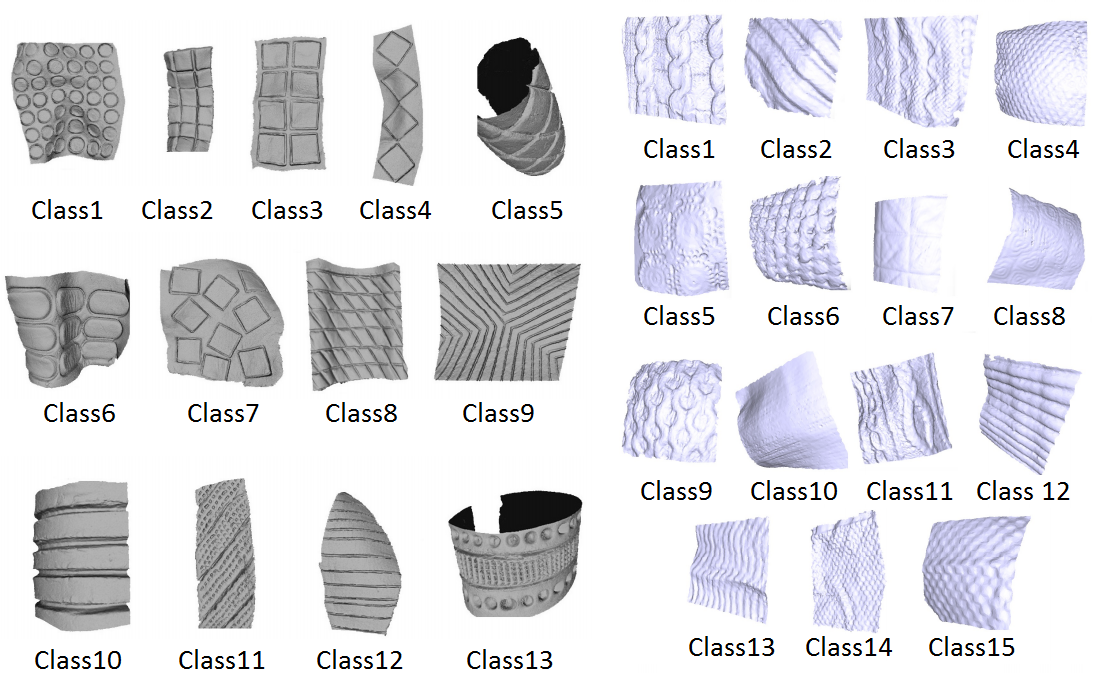}
\begin{tabular}{p{2cm}p{3cm}p{2.5cm}p{3cm}}
 &(a)& & (b)
\end{tabular}
\caption{(a): the 13 models used to originate the first dataset. (b): the knitted patterns of the SHREC17 contest \cite{shrec17}.}
\label{fig:dataset_e}
\end{figure}
We adopt two datasets.
\begin{itemize}
\item \emph{Plastic Dataset}: it is composed by 52 triangle meshes derived from the laser scans of 13 physical models, each one representing a specific geometric pattern (e.g., lines, circles, squares incised on a surface, see Figure \ref{fig:dataset_e}a). For each model, we get two separate surfaces for a total of 26 meshes, which together form the \emph{Original Dataset}. Then, for each surface we simulate erosion and degradation with a Laplacian smoothing filter, for a total of 52 meshes (4 variations for each pattern), which together form the \emph{Complete Dataset}. The whole dataset is freely available\footnote{The dataset is available at: https://github.com/EliaMTH/das-data}.
\item \emph{SHREC'17 dataset}: it corresponds to the recent SHREC'17 benchmark \cite{shrec17} on the retrieval of relief patterns. It is composed by 720 triangle meshes derived from knitted objects, grouped into 15 classes (see Figure \ref{fig:dataset_e}(b)), each one made of 48 textile patterns. Each class has been created from 15 base surfaces (embedding a single textile pattern into 12 different positions); then, each surface was modified with four mesh re-samplings.
Again, two datasets can be derived: the first one is related to the complete dataset of 720 models and aims at evaluating the overall robustness and stability of methods with respect to different mesh representations. The second one groups the 180 original meshes according to their textile pattern and it is better suited to analyze the capability of a method of effectively recognizing a pattern independently of the overall surface embeddings. 
\end{itemize}

\subsection{Evaluation measures}
\label{sec:evalperformances}
The evaluation tests have been performed using a number of classical information retrieval measures, namely the Nearest Neighbor, First Tier, Second Tier, Discounted Cumulative Gain, e-measure, Precision-Recall plot, confusion matrices and tier images.

\paragraph*{Nearest Neighbor, First Tier, Second Tier}
These measures aim at checking the fraction of models in the query's class also appearing within the top $k$ retrievals. 
In detail, for a class with $|C|$ members, $k=1$ for the Nearest Neighbor (NN), $k =|C|-1$ for the first tier (FT), and $k = 2(|C| - 1)$ for the second tier (ST). Note that all these values range from 0 to 1.

\paragraph*{Discounted cumulative gains}
The Discounted Cumulative Gain (DCG) is an enhanced variation of the Cumulative Gain, which is the sum of the graded relevance values of all results in the list of retrieved objects of a given query. 
%
%
%
\paragraph*{Precision-Recall and mAP}
The \emph{Precision} and \emph{Recall} are two common measures for evaluating search strategies. 
Recall is the ratio of the number of relevant records retrieved to the total number of relevant records in the database, while precision is the ratio of the number of relevant records retrieved to the size of the return vector \cite{salton_evaluation}. 
Precision and recall always range from 0 to 1. Often a visual interpretation of these quantities is plot as a curve in the reference frame recall vs. precision \cite{Baeza-Yates:1999}: the larger the area below such a curve, the better the performance under examination.
As a compact index of precision vs. recall, we consider also the mean Average Precision (mAP), which is the portion of area under a precision-recall curve. 

\paragraph*{e-measure}
Since the mostly interesting query results are the first ones retrieved, the e-measure $e$~\cite{Rijsbergen1979} was introduced as a quality measure of the first models retrieved for every query. The $e$ measure depends on the \emph{Precision} and \emph{Recall} values by the relation: $$e=\frac{2}{Precision^{-1}+Recall^{-1}}.$$

\paragraph*{Confusion matrices and Tier images}
Each classification performance can be associated with a confusion matrix $CM$, that is, a square matrix whose dimension is equal to the number of classes in the dataset. For the row $i$ in $CM$, the element $CM(i,i)$ gives the number of items which have been correctly classified as elements of the class $i$; similarly, elements $CM(i,j)$, with $j\neq i$, count the items which have been misclassified, resulting as elements of the class $j$ rather than elements of the class $i$. Thus, the classification matrix $CM$ of an ideal classification system should be a diagonal matrix, such that the element $CM(i,i)$ equals the number of items belonging to the class $i$.
Similarly, the tier image visualizes the matches of the NN, FT and ST. The value of its element $(i,j)$ is: \emph{black} if $j$ is the NN of $i$, \emph{red} if $j$ is among the $(|C|-1)$ top matches (FT) and \emph{blue} if $j$ is among the $2\cdot (|C|-1)$ top matches (ST). When the models of a class are grouped along each axis, the optimal tier image clusters the black/red square pixels on the diagonal.
\section{edgeLBP properties and performances}
\label{sec:ex_results}
Being defined on the basis of shape-based criteria and a region-growing approach that visits the surface edges that are closer than $R$ to a vertex, the edgeLBP is naturally rotation-invariant. 
Since we are interested in local geometric variations, the radius of the spheres must be kept quite small. Keeping in mind that a Riemannian surface can be locally approximated as a Euclidean space \cite{Bronstein:2008} (in our case a disk) the edgeLBP descriptor is robust to different surface bendings. 
Moreover, the edgeLBP is able to characterize the surface of generic 3D models and it is insensitive to object obstructions.

Overall, we evaluated the edgeLBP description over thousands of models. From that analysis, we selected the examples the most representatives of the edgeLBP properties. 
In the reminder of this Section, we discuss the robustness of the edgeLBP with respect to different mesh re-samplings and decimations,  vertex perturbations with Gaussian noise and the type of faces used in the tessellation (Sections \ref{sec:rob_stab}, \ref{sec:noise} and \ref{sec:quad}, respectively).
We validate our approach on the plastic dataset and compare its performance against the meshLBP description in Section \ref{sec:analysis}.

\subsection{Robustness with respect to different surface samplings}

\label{sec:rob_stab}
\begin{figure}[ht]
\centering
\begin{tabular}{cccc}
Model 1& Model 2&Model 3&Model 4\\
\multicolumn{4}{c}{\includegraphics[width=11cm]{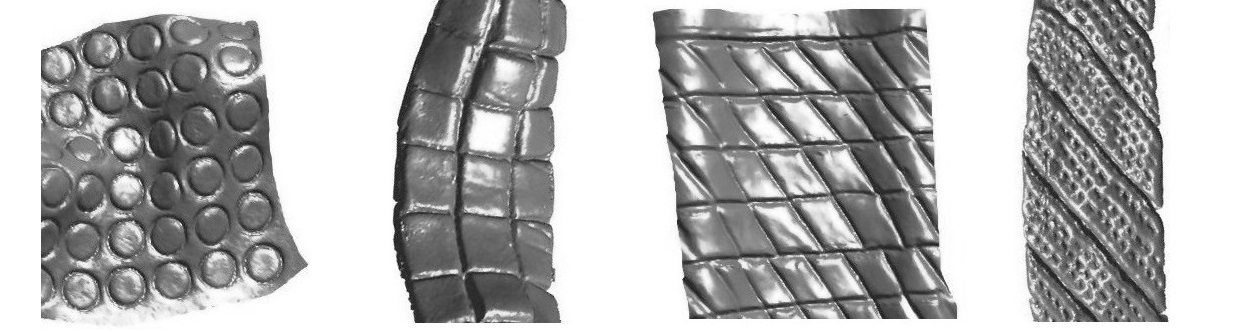}}\\
\includegraphics[angle=90,width=2.5cm]{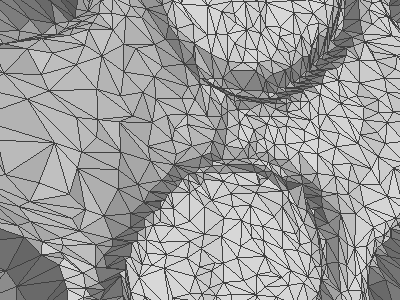}&
\includegraphics[angle=90,width=2.5cm]{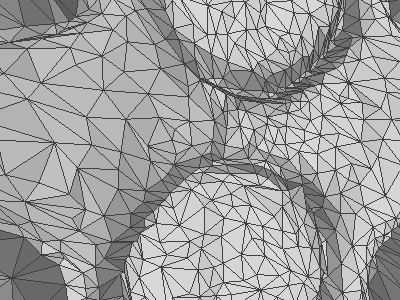}&
\includegraphics[angle=90,width=2.5cm]{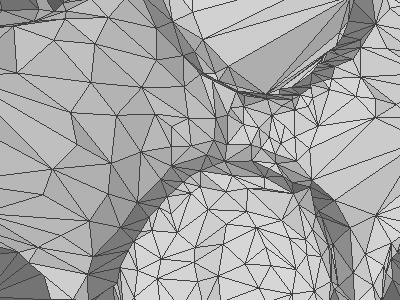}&
\includegraphics[angle=90,width=2.5cm]{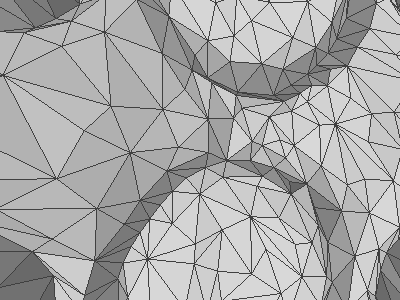}\\
$20K$ & $15K$ & $10K*$ & $5K$\\
\end{tabular}
\caption{First row: Four surface models from the plastic dataset. Second row: Details on the re-samplings of the model 1; $10K*$ highlights the local effect of the manual non-uniform sampling.}
\label{fig:model_confronto}
\end{figure}

To test the robustness of the edgeLBP description against different mesh samplings we consider four models that represent four types of possible geometric patterns, see Figure \ref{fig:model_confronto}.
\begin{figure}[t]
\centering
\begin{tabular}{ccc}
meshLBP& edgeLBP\\
\includegraphics[width=5.5cm]{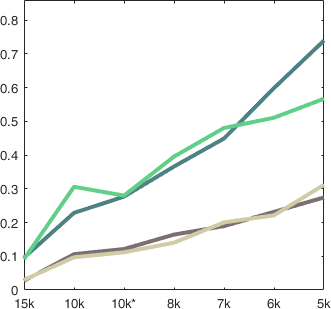} &
\includegraphics[width=5.5cm]{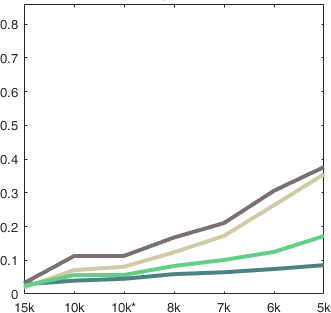}\\
\multicolumn{2}{c}{\includegraphics[scale=1]{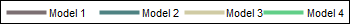}}\\
meshLBP $ring_1$ examples & edgeLBP $ring_1$ examples\\
\includegraphics[width=5cm]{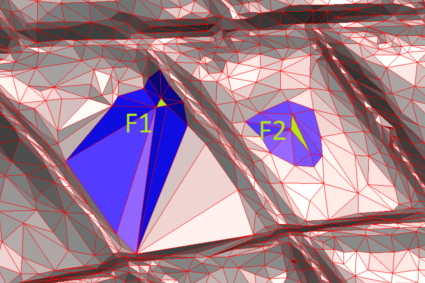}&
\includegraphics[width=5cm]{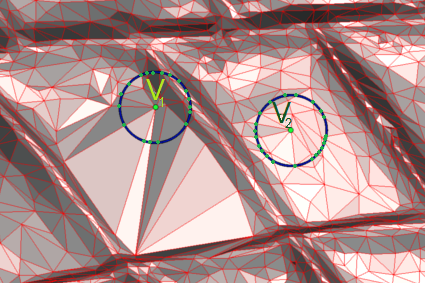}
\end{tabular}
\caption{Top: graphical plot of the distances from the original model when the number of samplings diminishes, computed with respect to the meshLBP and the edgeLBP descriptions. The x-value $10K*$ represents the mesh with 10000 vertices manually edited. Bottom: representation of $ring_1$ in both algorithms.}
\label{fig:macro_confronto}
\end{figure}
Then, we simplified these models removing vertices from the mesh according to a minimum geometric approximation error criterion. From the original mesh we derived a sequence of meshes that best approximates it with a fixed number of vertices. Meshes are automatically created using the ReMesh tool \cite{remesh}, with $15K$, $10K$, $8K$, $7K$, $6K$, $5K$ vertices. For each model, one more mesh was manually edited in order to have an un-even sampling with vertices concentrated in some parts of the surface (see Figure \ref{fig:model_confronto}, mesh $10K*$).

Figure \ref{fig:macro_confronto}(Top) represents the $\chi^2$ distance between the edgeLBP descriptor of the original mesh (the one with $20k$ vertices) and its variations. For each test, the distance values are normalized with respect the maximum value of the $\chi^2$ distance. The x-axis represents the number of vertices of the mesh, while the y-axis represents the distance from the original model. 
We repeat the same analysis for the meshLBP description.
We run both methods with two different parameter settings: twelve samples and seven rings (\emph{run1}) ($P=12$ and $N_r=7$ in our notation, which corresponds to the standard configuration of the meshLBP, as released in the Matlab toolbox\footnote{https://it.mathworks.com/matlabcentral/fileexchange/48875-mesh-lbp}) and $P=15$ and $N_r=5$ (\emph{run2}). 
As the geometric property for these experiments we select the maximum curvature, $k_2$; the weight function $\alpha_1$ is used for both edgeLBP and meshLBP. Since the results are qualitatively the same for both runs, Figure \ref{fig:macro_confronto} reports the outcome of the \emph{run2}, only.

Not surprisingly, the edgeLBP descriptor shows a stronger stability towards the mesh decimation and corruption when compared with the outcome of the meshLBP description. We think that this effect is mainly due the sensitivity to the mesh tessellation of the face-based expansion method: Figure \ref{fig:macro_confronto}(Bottom) shows the $ring_1$ for both the edgeLBP and the meshLBP descriptions on two vertices of the same mesh. Indeed the sphere-surface intersection is robust to mesh irregularities and permits the LBP to cope a wide variety of tessellations.

%
\subsection{Robustness with respect to noise}
\label{sec:noise}

To evaluate how the edgeLBP depends on the data quality, the vertices of the meshes of the Plastic dataset have been perturbed with geometric noise, by modifying the vertex coordinates.
Since we are interested in geometric patterns that represent surface reliefs and/or chiseled decorations we must keep a reasonable balance between the intensity of the vertex perturbation and the size of the pattern. Indeed, the presence of noise can significantly alter the nature of the pattern and it is recognized as one of the open challenges for 3D pattern recognition \cite{shrec_geo}.
To this end, we simulate a geometric perturbation of the mesh modifying the coordinates of the vertices through a random Gaussian perturbation applied at four levels of intensity. The variance of the perturbation ranges from 0.4\% to 1.6\% of the maximum diameter of the model. Note that, the intensity of the vertex perturbation, depending on the model diameter, varies from model to model.

\begin{figure}
\centering

\begin{tabular}{ccccc}
Original & $0.4$\% &$0.8$\% & $1.2$\% & $1.6$\%\\ \includegraphics[width=2.3cm]{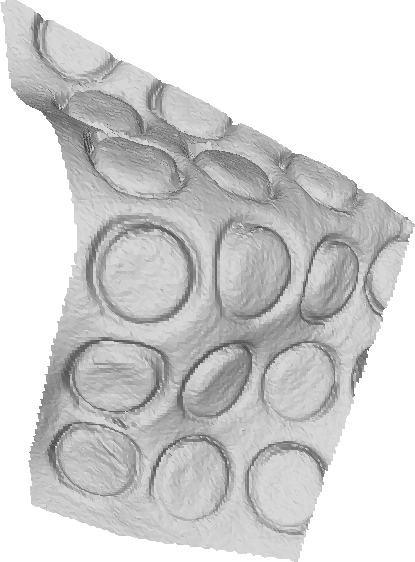}&
\includegraphics[width=2cm]{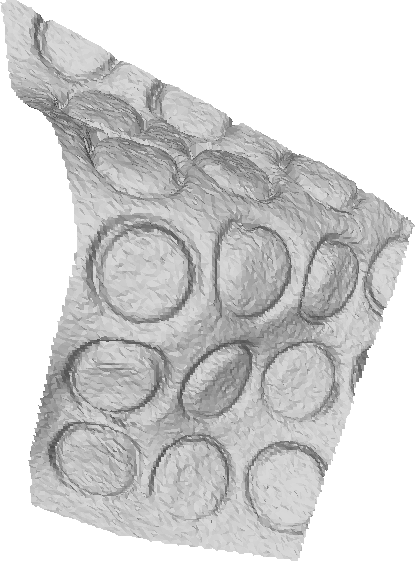}&
\includegraphics[width=2cm]{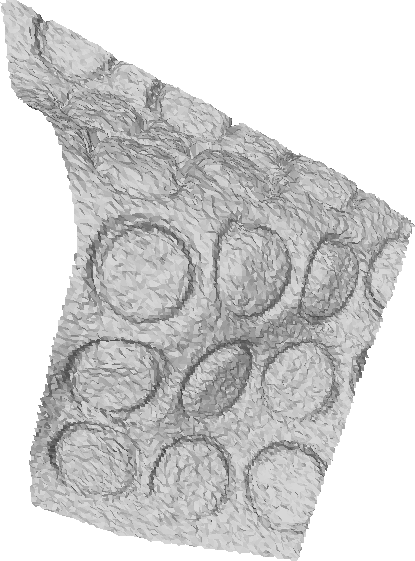}&
\includegraphics[width=2cm]{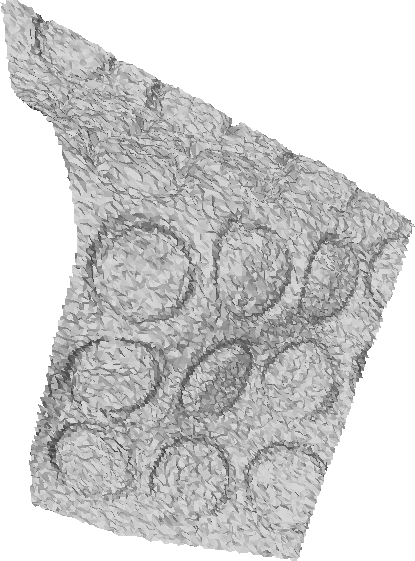}&
\includegraphics[width=2cm]{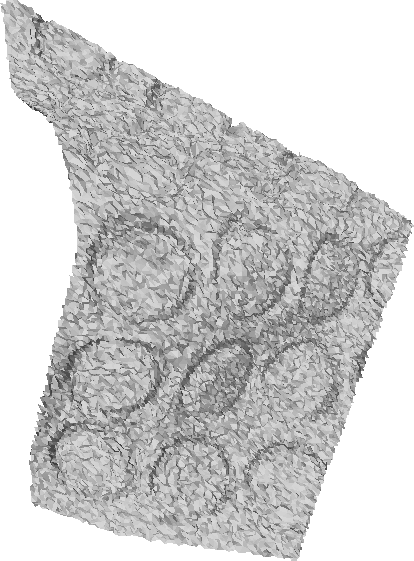}
\end{tabular}
\small
\begin{tabular}{|c|c|c|c|c|c|c|}
\hline
	   		&NN 	& 1-Tier & 2-Tier  & mAP  &e   & DCG 	\\
\hline
Original models		&0.87&0.87&0.99&0.82& 0.17&0.81\\
\hline
$noise=0.4$\%		&0.85&0.89&0.99&0.82 &0.16&0.81\\
\hline
$noise=0.8$\%		&0.84&0.90&0.99&0.83&0.16&0.81\\
\hline
$noise=0.12$\%		&0.63&0.75&0.97&0.78& 0.16&0.76\\
\hline
$noise=0.16$\%		&0.69&0.77&0.93&0.78& 0.16 &0.76\\
\hline
\end{tabular}
\caption{Top: examples of the noise added to the model of the Complete Plastic Dataset.
Bottom: The evaluation measures of the runs, compared to the performances on the clean models.}
\label{pic:noise_results}
\end{figure}

The top row of Figure \ref{pic:noise_results} shows the increasing intensity of the vertex perturbation over a model of the dataset. Overall, Figure \ref{pic:noise_results} highlights that the edgeLBP is robust with respect to perturbations of the vertex coordinates.

\subsection{Robustness with respect to different surface representations}
\label{sec:quad}
\begin{figure}[t]
\centering
\begin{tabular}{cc}
\includegraphics[width=5cm]{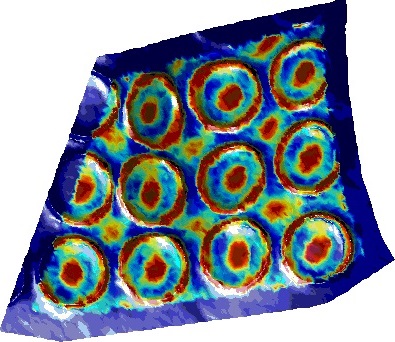}&
\includegraphics[width=5cm]{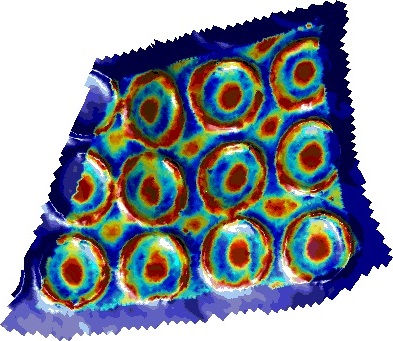}\\
Triangle mesh & Quad mesh\\
\end{tabular}
\caption{EdgeLBP values using two tessellations of the same sur
face. The values of the edgeLBP on the two tessellations are represented with the same jet color-map. Blue vertices close to surface boundary represent the non-admissible ones.}
\label{fig:quad_tr}
\end{figure}

As a further contribution to the panorama of local feature descriptors, our definition of the edgeLBP is able to deal with polygonal tessellations, not only triangulations (on the contrary of the meshLBP that is strongly based on the triangle-mesh structure). 
Among the experiments we conducted to assess the coherence of the edgeLBP description across different types of tessellation, Figure \ref{fig:quad_tr} represents the edgeLBP values on two different tessellations of the same surface, namely a triangle and quadrangle mesh. 
%
%
%
Colors are used to represent the values of the edgeLBP on the mesh vertices; same color  corresponds to the same value of the edgeLBP. As expected, the color distribution on the surfaces is the same. This means that the edgeLBP on both tessellations assumes the same values over the vertices.

%

Besides tessellations, the edgeLBP description can be used to analyze also voxel grids and point clouds. However, an ad-hoc extension of the LBP to voxel grids already exists \cite{3DLBPvoxel} and the use of a sphere-grid intersection is computationally redundant. On the contrary, we think that the edgeLBP description can successfully deal with point clouds, for instance, adopting a representation based on the kdtree structure and following the implementation strategy adopted in the Point Cloud Library (PCL\footnote{http://pointclouds.org/}) for several feature descriptors.
\subsection{Performances on the plastic dataset}
\label{sec:analysis}

In this Section, we compare the retrieval and classification performances of the edgeLBP with the meshLBP, the FPFH \cite{Rusu:2009}, the SHOT~\cite{TombariSS11}, the Spin Images (SI) \cite{Johnson:1999} on the complete Plastic dataset. 
The implementations of the SI and FPFH descriptors come from the Point Cloud Library, while for the SHOT description we adopt the authors' implementation available on GitHub\footnote{https://github.com/fedassa/SHOT}. Both these libraries come with default parameters settings and we use those for our runs.

For the edgeLBP and meshLBP, we used $k_2$ as $h$ function and the weight function $\alpha_1$. We adopt the two configurations \emph{Run1} and \emph{Run2} of the the parameters  described in Section \ref{sec:rob_stab}; finally, for the edgeLBP $R_{max}=2.5mm$.

\begin{figure}[t]
\centering
\begin{tabular}{ccc}
$meshLBP$ & $edgeLBP - \emph{Run1}$ & $edgeLBP - \emph{Run2}$ \\
\includegraphics[width=3.2cm]{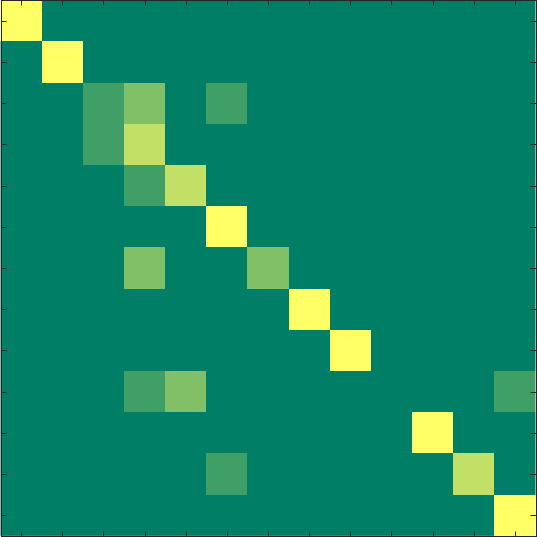}&
\includegraphics[width=3.2cm]{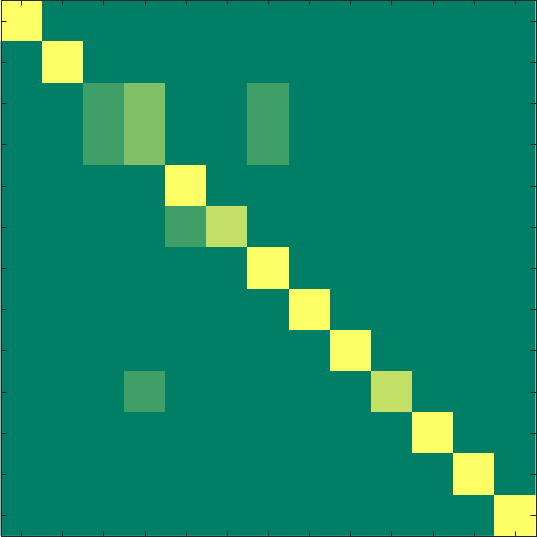}&
\includegraphics[width=3.2cm]{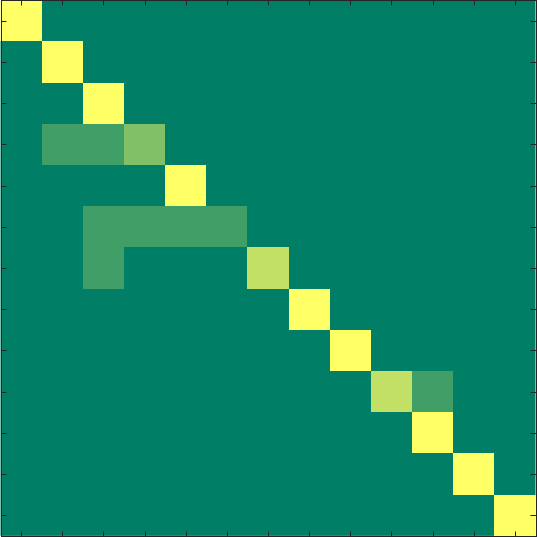}\\
$Spin Images$ & $SHOT$ & $FPFH$ \\
\includegraphics[width=3.2cm]{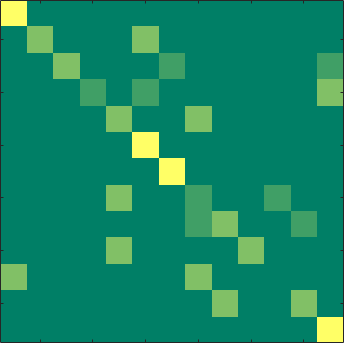}&
\includegraphics[width=3.2cm]{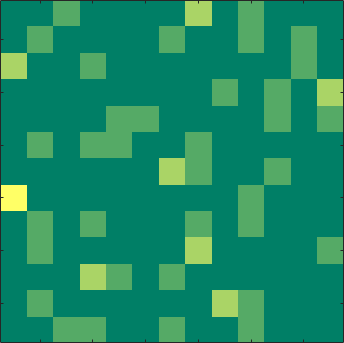}&
\includegraphics[width=3.2cm]{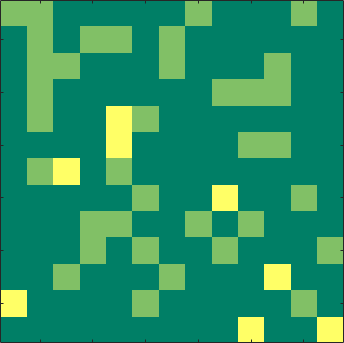}
\end{tabular}
\caption{Confusion matrices of the runs over the plastic dataset.}
\label{pic:all_methods}
\end{figure}

Table \ref{tab:plastic_dataset} reports the retrieval and classification performances of the runs considered. For each run, the left column lists the best feature vector distance. Figure \ref{pic:all_methods} presents the confusion matrices of each run listed in Table \ref{tab:plastic_dataset}.
\begin{table}
\caption{Retrieval and classification performances over the plastic dataset.  B. abbreviates the term Bhattacharyya.}
\label{tab:plastic_dataset}
\small
\begin{tabular}{|c|c|c|c|c|c|c|}
\hline
	Method   		&NN 	& 1-Tier & 2-Tier  & mAP    & e & DCG 	\\
\hline
meshLBP	(\emph{B.} distance)		&$0.77$ & $0.74$ & $0.86$  & $0.73$ & 0.17 & $0.71$\\
\hline
edgeLBP - \emph{Run1} (\emph{B.} distance)		&$0.87$ & $0.82$ & $0.99$  & $0.82$ & 0.17 & $0.80$\\
\hline
edgeLBP - \emph{Run2} (\emph{B.} distance)		&$0.87$ & $0.87$ & $0.99$  & $0.82$ & 0.17 & $0.81$\\
\hline
Spin Images (\emph{B.} distance) &$0.58$& $0.55$  &  $0.70$ & $0.67$ & $0.16$ & $0.65$\\
\hline
SHOT ($\chi^2$ distance)		&$0.23$& $0.21$  & $0.23$  & $0.34$      & $0.09$ & $0.31$\\
\hline
FPFH (\emph{EMD} distance)		&$0.23$& $0.23$  & $0.32$  & $0.43$ & $0.1$ & $0.39$ \\
\hline
\end{tabular}
\end{table}

As expected, the edgeLBP and the meshLBP outperform the other local feature descriptions that, as discussed in Section \ref{sec:star}, are better tailored for shape matching and registration rather than for the comparison of the pattern reliefs. In particular, the SHOT and the FPFH descriptions often confuse incised or relief features like circles, straight lines and squares, probably because the overall distribution of the feature points is compatible but there is not a spatial encoding of their relative position. 
Indeed, SHOT and FPFH detect sets of feature points on each model and map the models on the basis of their points correspondence. On the other hand, Spin Images are designed for partial matching. While these two strategies are well suited for partial matching and registration, they results limited for the comparison of repeated surface reliefs that are repeated on the surfaces a different number of times.
%
%
%
%
\section{SHREC'17 pattern retrieval contest dataset}
\label{sec:shrec}
To compare the edgeLBP with the other participants to the SHREC'17 pattern retrieval benchmark we use the two configuration parameters (\emph{run1} and \emph{run2}) proposed in Section \ref{sec:rob_stab} and \ref{sec:analysis}. $R_{max}$ is set $10mm$ for both runs and is obtained by measuring the size of the patterns of three, randomly selected, surfaces. 
	\begin{table}[t]
		\centering
		\caption{Retrieval Performances obtained by the edgeLBP compared with the scores of the best performing methods in \cite{shrec17} on the SHREC17 datasets. The best runs are in bold.}
		\scriptsize{
			\begin{tabular}{c}
				\begin{tabular}{|c|c|c|c|c|c|c|}
					\hline
					\multicolumn{7}{|c|}{\normalsize{Original Dataset}}\\
					\hline
					Method  & NN       & 1-Tier  & 2-Tier & mAP    &  e	  & DCG   \\ \hline
					LBPI         & 0.339    & 0.207   & 0.353  & 0.250  & 0.237 & 0.250 \\ \hline
					GI HOG       & 0.089    & 0.069   & 0.130  & 0.118  & 0.097 & 0.373 \\ \hline
					IDAH-2       & 0.339          & 0.182          & 0.271           & 0.215          & 0.181          & 0.503         \\ \hline
					CMC-1        & 0.600    & 0.342   & 0.461  & 0.371  & 0.274 & 0.641 \\ \hline
					CMC-2        & 0.633	  & 0.363 	& 0.494  & 0.390  & 0.293 & 0.662 \\ \hline
					CMC-3        & 0.533    & 0.281   & 0.394  & 0.308  & 0.242 & 0.596 \\ \hline
					SQFD-HKS     & 0.106          & 0.066          & 0.137          & 0.123          & 0.102          & 0.376          \\ \hline
					KLBO-FV-IWKS & 0.522    & 0.295   & 0.412  & 0.307  & 0.247 & 0.603 \\ \hline
					KLBO-SV-IWKS & 0.489    & 0.249   & 0.375  & 0.273  & 0.235 & 0.570 \\ \hline  edgeLBP - \emph{run1}      & \textbf{0.922}    & 0.683   & 0.825 & 0.716  & 0.580 & 0.863 \\ 
\hline  edgeLBP - \emph{run2}     & 0.911    & \textbf{0.689}   & \textbf{0.844}  & \textbf{0.725}  & \textbf{0.590}  & \textbf{0.865} \\ \hline
\end{tabular}
\\
\\
\begin{tabular}{|c|c|c|c|c|c|c|}
\hline
\multicolumn{7}{|c|}{\normalsize{Complete Dataset}}\\
\hline
Method                   & NN             & 1-Tier        & 2-Tier    & mAP	 &	e  	 & DCG	\\ \hline
LBPI         & 0.828          & 0.248         & 0.400     & 0.283  & 0.232 & 0.697	 \\ \hline
GI HOG       & 0.686          & 0.107         & 0.176     & 0.131  & 0.102 & 0.561 \\ \hline
IDAH-2       & 0.306          & 0.141         & 0.244     & 0.163  & 0.127 & 0.559\\ \hline
CMC-1        & 0.718          & 0.258         & 0.372     & 0.260  & 0.247 & 0.673 \\ \hline
CMC-2        & 0.763          & 0.272         & 0.389     & 0.271  & 0.261 & 0.686 \\ \hline
CMC-3        & 0.647          & 0.219         & 0.323     & 0.218  & 0.208 & 0.639 \\ \hline
SQFD-HKS     & 0.536          & 0.117         & 0.192     & 0.139  & 0.110 & 0.558 \\ \hline
KLBO-FV-IWKS & \textbf{0.986} 		  & 0.333 		  & 0.449 	  & 0.339  & 0.332 & 0.759 \\ \hline
KLBO-SV-IWKS & 0.978          & 0.287         & 0.409     & 0.296  & 0.283 & 0.732 \\ \hline  edgeLBP - \emph{run1}     & 0.979    & 0.619   & 0.763  & 0.651 & 0.413 &  0.894 \\ \hline  
edgeLBP - \emph{run2}     & \textbf{0.986}    & \textbf{0.634}   & \textbf{0.780}  & \textbf{0.669} & \textbf{0.421} &  \textbf{0.902} \\ \hline
\end{tabular}
\end{tabular}
}
\label{fig:shr_comparefull}
\end{table}
\begin{figure}[t]
\centering
\begin{tabular}{cccc}
CMC-2 & KLBO-FV-IWKS & LBPI & edgeLBP - \emph{run2}\\
\includegraphics[width=2.6cm]{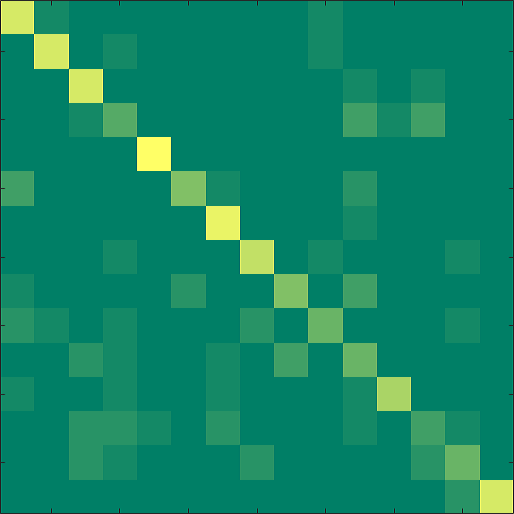}&
\includegraphics[width=2.6cm]{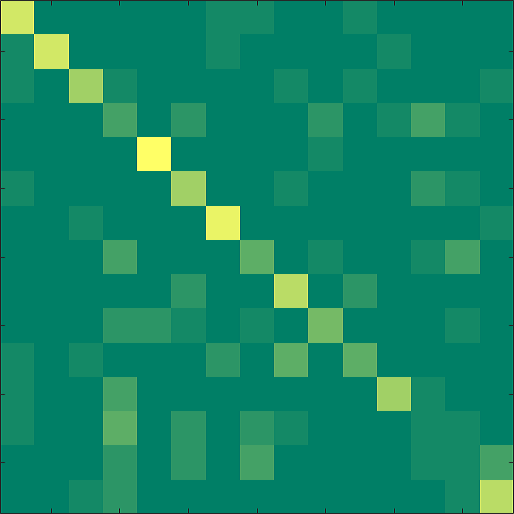}&
\includegraphics[width=2.6cm]{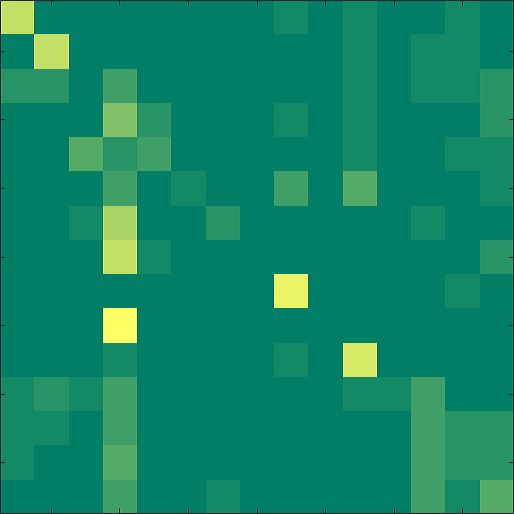}&
\includegraphics[width=2.6cm]{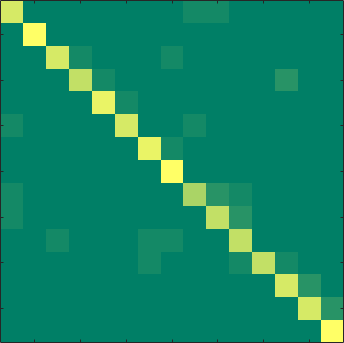}\\
\end{tabular}
\caption{Best confusion matrices in the SHREC17 retrieval pattern contest (Original Dataset) in comparison to that of the edgeLBP.}
\label{fig:cm_shrec}
\end{figure}

Table \ref{fig:shr_comparefull} reports the edgeLBP performances on the Original and Complete datasets and the best runs obtained by the SHREC'17 participants, who are indicated with the same label used in the SHREC'17 report \cite{shrec17}.
In the case of the Original Dataset, our method provides the best results in all the scores, showing a good capability of discriminating the geometric patterns.
In the case of the complete dataset, our results significantly overcome the other participants in each measure, only the NN measure is equivalent to the (KLBO-*) runs.  As stated in the SHREC report \cite{shrec17}, the NN value of methods that analyze the global geometry (such as the KLBO-* run) is biased by the presence of three variations of each mesh in the original dataset. In this case, each mesh sampling keeps the same overall embedding but degrades the mesh connectivity and approximates the original reliefs. For this reason, global methods are made easy to find as the Nearest Neighbor one of the mesh variations; however, they rapidly degrade when the query range increases, like reflected by the FT and ST scores.

\begin{figure}[t]
\centering
\begin{tabular}{cc}
Original Dataset& Complete Dataset\\
\includegraphics[width=5.5cm]{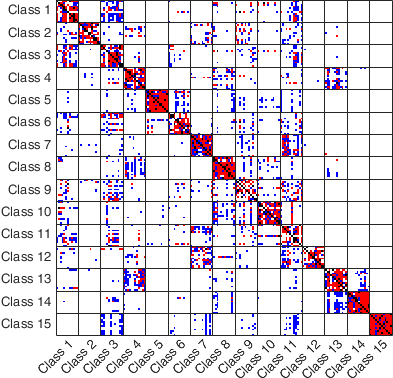}&
\includegraphics[width=5.5cm]{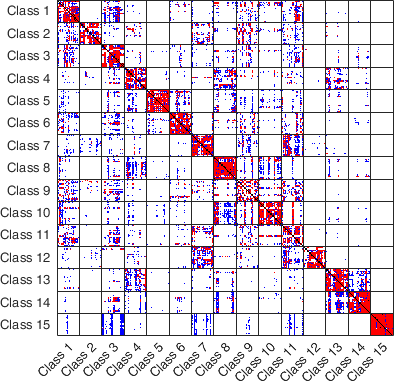}\\
&\\
\includegraphics[width=5.5cm]{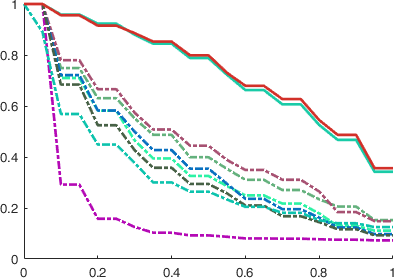}&
\includegraphics[width=5.5cm]{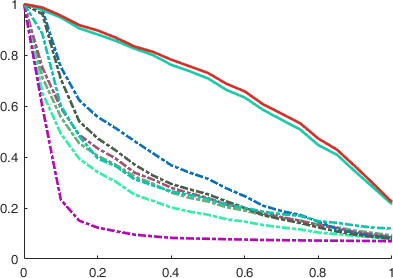}\\
\multicolumn{2}{c}{\includegraphics[width=6cm]{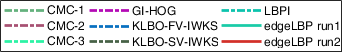}}\\
\end{tabular}
\caption{Tier image and Precision-Recall plots of the original dataset (left) and complete dataset (right), using the edgeLBP - \emph{run2}. In the tier images, rows represent the queries, the NN is marked in black, the FT in red and the ST in blue.}
\label{fig:conf_tier}
\end{figure}

Figure \ref{fig:cm_shrec} compares the confusion matrix derived from the best edgeLBP run (\emph{run2}) against the three best runs in \cite{shrec17}. Similarly, Figure \ref{fig:conf_tier}(top) depicts the tier images both on the Original and the Complete datasets of the edgeLBP, \emph{run2}. In the case of the complete dataset, it is worth noticing that adding three variations of the original patches does not alter the overall mask of the tier image, highlighting the coherence of the retrieval performance among the original dataset and its variations.

Finally, Figure \ref{fig:conf_tier} (bottom) plots the Precision-Recall curves in both the dataset configurations and compares the best edgeLBP performance (\emph{run2}) with the best runs in \cite{shrec17}.
The overall performance of the edgeLBP successfully deals with the SHREC'17 dataset and generally improves with respect to the other participants of more than $20\%$; in our opinion this reveals that the method combines the efficacy of a local pattern characterization with a mesh independent and embedding invariant description. 
\subsection{Analysis of the edgeLBP parameters}
\label{sec:settings}
We now discuss the performance of the edgeLBP description when different parameter settings are chosen. For the simplicity of its storage, all these experiments are performed using the weight function $\alpha_1$.

Besides the $k_2$ function adopted in the experiments exhibited in the previous Sections, we tested all the geometric functions described in Section \ref{sec:properties}. In these experiments we fixed $P=12$, $N_r=7$ and $R=10mm$.
Table \ref{tab:parameters_tables}(left) compares the retrieval performance of the edgeLBP (in terms NN, FT and ST) on the Original dataset of the SHREC'17 benchmark when the geometric functions vary. 
\begin{table*}
\centering
\scriptsize
\caption{Top: Retrieval performance of the edgeLBP for different geometric functions. Bottom: Retrieval performance when the edgeLBP parameter settings vary. In the runs (*) and (**) the models are simplified at 10K and 11K vertices, respectively. All tests are performed on the Original Dataset of the SHREC'17 benchmark. The number between $< >$ brackets is the size of the feature vector.}
\begin{tabular}{c}
\\
\begin{tabular}{|c|c|c|c|}
	\hline
		Curvature ($h$)		&	NN	&	FT	&	ST	\\
    \hline
	$k_1$ 	& 0.88	&	0.65&	0.80 \\
    \hline
    $k_2$	& 0.92  &	0.68&	0.84 \\
    \hline
    $K$		& 0.87 	&	0.61&	0.78 \\
    \hline
    $H$		& 0.87	& 	0.61&	0.78 \\
    \hline
    $SI$	& 0.66 	&	0.44&	0.59 \\
    \hline
    $Curvedness$ & 0.71 & 0.48 & 0.63\\
    \hline
\end{tabular}
\\
\\
\begin{tabular}{|c|c|c|c|}
	\hline
			edgeLBP Parameters	&	NN	&	FT	&	ST	\\
    \hline 
    $N_r=7$, $P=12$, $R_{max}=10mm$, $<84>$ (*)& 0.85	&	0.62&	0.78 \\
    \hline
    $N_=7$, $P=12$, $R_{max}=14mm$, $<84>$ 	(**)& 0.88  &	0.66&	0.82 \\
    \hline
     $N_r=5$, $P=12$, $R_{max}=5mm$, $<60>$ 	& 0.83	&	0.59& 	0.72 \\
	\hline
     $N_r=5$, $P=8$, $R_{max}=10mm$, $<40>$ 	& 0.86	& 	0.65& 	0.80 \\
	\hline
    $N_r=5$, $P=18$, $R_{max}=10mm$, $<90>$ 	& 0.89	&	0.68&	0.84 \\
	\hline
    $N_r=5$, $P=15$, $R_{max}=10mm$, $<75>$ 	& 0.91	&	0.69&	0.84 \\
    \hline
\end{tabular}
\end{tabular}
\label{tab:parameters_tables}
\end{table*}
From our experiments we noticed that $k_2$ performs slightly better than the other curvature-based properties. The fact that many curvature-based functions gave such high results is another confirmation of the descriptive power of the edgeLBP approach.

Regarding the other edgeLBP parameters we tested $N_r$ ranging from $3$ to $11$, $P$ was sampled between $8$ and $20$ and $R_{max}$ was varied from $5mm$ to $14mm$. 
As a summary of these comparison we report in Table \ref{tab:parameters_tables}(right) some NN, FT and ST performances. 
Notice that in this test we are considering also the number of the vertices $n_v$ of the tessellation as a parameter. In this case, we adopt an adaptive re-sampling to remove with higher priority the vertices that introduce a smaller approximation error of the shape. The smallest $n_v$, the roughest the approximation of the surface; when $n_v$ becomes very low (generally less than $10K$ vertices), the patterns on the surfaces considerably degrade and the number of the vertices does not kept small, geometric variations. In the other tests, we re-sampled all the meshes to $15k$ vertexes.
At the end of our analysis, the best edgeLBP parameter settings for the SHREC'17 benchmark are $h=k_2$, $P=15$, $N_r=5$, $R_{max}=10mm$ (the \emph{run2} in the previous Sections).
\section{Computational performances}
\label{sec:realcosts}
A prototype of the edgeLBP algorithm is implemented in MATLAB. Tests were computed on an Intel Core i7 processor (at 4.2 GHz). The main contribution to computational time comes from the number of vertices of the tessellation. The radial and spatial resolution marginally impact the final computational time, as far as the maximum radius is kept constant.
For a surface of roughly 4200 $mm^2$ represented by a triangle mesh with $n_v=20K$ vertices, the computation time of the edgeLBP operator with settings $R_{max}=2.5 mm$, $P=12$, $N_r=5$) is approximately 2 minutes. The computational time drops significantly if the vertices are reduced: the edgeLBP computation on the same surface with $n_v=10K$ samples ends approximately in 40 seconds and 25 seconds when $n_v \leq 8K$.
Overall, the edgeLBP computation on the whole SHREC'17 benchmark using the parameter settings presented in Section \ref{sec:settings} takes almost 18 hours.

The computational time of comparing two edgeLBP descriptors is in the order of $0.1 \cdot 10^{-5}$ seconds; for the SHREC'17 dataset (720 models) the computation of the dissimilarity matrix takes 2.5 seconds.
\section{Discussions and conclusive remarks}
\label{sec:conclusion}
We have extended the LBP concept to surfaces and defined a novel description, whose core strength is its independence from the mesh tessellation. The method provides an effective coding of the local shape properties and the experimental results show its efficacy in the detection of patterns on surfaces.
The edgeLBP description is invariant to roto-translations. 
With respect to the previous LBP extensions to depth surfaces, our description does not require any model normalization, registration or projection with depth maps. Experiments show that the edgeLBP is independent of self-occlusions and it is able to handle mesh data with or without boundary. If compared with the meshLBP, the edgeLBP is robust to mesh simplification, does not require any uniform mesh re-sampling nor a specific type of mesh faces (triangles). 

Having the edgeLBP independent of the physical size of the patterns depends on the application. On one hand, scale invariance is fundamental if the task is to retrieve all the diamond-like patterns in a dataset. In our method, scale and affinity invariance are obtained selecting the $R_{max}$ value as a fraction of the square-root of the surface area. On the other hand, the size of the geometric pattern can be an important aspect like in the datasets considered in our experiments because the objective is to retrieve patterns with the same physical appearance: for this reason, on these experiments, we selected a unique $R_{max}$ value for each dataset.
At the light of this reasoning, we briefly discuss the performance of the edgeLBP with respect to the type of patterns considered in this patters (representative pictures of the class elements are shown in Figure \ref{fig:dataset_e}). 
\paragraph*{Plastic dataset} The elements of the Class 6 are characterized by elongated, incised ovals; not surprisingly they are somehow confused with incised squares and diamond-like incisions (Classes 3 and 5) that present a similar distribution on the surface and physical, comparable size. Again, small patterns are well characterized (Classes 1, 8, 11, 13). An interesting relation is the one between the Classes 1 and 8: the Class 8 is the closest to the Class 1 (because of the small size of circles and diamonds) but the classes the closest to the Class 8 are the Classes 1 and 5 (the Class 5 is another pattern with diamonds).
Moreover, we notice that an important factor of the method is its sensitiveness to the global distribution of a pattern on the surface as highlighted by the similarity between the Classes 3 and 7. 
However, the use of a single value for the $R_{max}$ parameter explains the confusion between elements of the Classes 10 and 11, because the strips in the Class 11 are the same of those in the Class 10 with the addition of small incisions on the top. 

\paragraph{SHREC'17 benchmark} Fragments with a pattern composed by class-specific features (with respect to those present in this dataset) are the easiest to recognize for the edgeLBP. Also, classes that are visually close (like the Classes 8 and 10) are correctly classified and judged as similar. The Class 4 was misclassified by all the methods that participated at the SHREC'17 contest. On the contrary, the edgeLBP is able to correctly classify (using NN) 47 patches over the 48.
The tier image reveals that the closest classes to the Class 4 are the Classes 8 and 13 that are defined by features having comparable geometric size.
The Class 6 is another challenging class: its elements were significantly corrupted by the mesh re-sampling operations and the bumps of this pattern are present with approximately the same geometric size in other classes. Nevertheless, the edgeLBP correctly classifies 47 patches over 48. 
Among all the elements of the dataset, the patterns of the Classes 1, 3 and 11 form a subset made of knitted fabrics like twists, diamonds, and so on. Despite the relevance of the relief in these decorative elements, their pattern elements are less repeated on the surface patch, for instance, the pattern element in the representative of the Class 1 in Figure \ref{fig:dataset_e}(c) is repeated 2.5 times. Over a dataset made of surfaces having approximately the same dimension (10 x 10 cm), this fact influences the edgeLBP performance because it has the alternate count of the feature occurrences at his core.
\paragraph*{Future plans}
Further extensions are planned and possible. For instance, it is possible to extend this approach to colorimetric patterns, using one color channel as the $h$ function.
Moreover, we are currently working on how to store multidimensional properties in the edgeLBP operator. This is the case of the color spaces that are treated by the edgeLBP representation one channel at a time.

In addition, while the sphere-based intersection determines rings that capture isotropic shape properties, we plan to follow the LBP approaches proposed in the literature of depth images to extend our rings to anisotropic ones, such as ellipses and other curve variations.

Finally, we think that this contribution paves the road towards the automatic recognition of multiple patterns on surfaces. Indeed, current experiments are performed on surfaces fully characterized by a single pattern at a time; next plans include the combination of the shape description step with segmentation techniques and the aggregation of parts made of vertices with similar local descriptions, for instance following an approach similar to \cite{Itskovich2011}.
\section*{Acknowledgments}
The authors thank Bianca Falcidieno and Michela Spagnuolo for their support and the helpful discussions and suggestions on this topic. The work is developed within the research programme of the ``H2020'' European project ``GRAVITATE'', contract n. 665155, (2015-2018).
%
%
%
%
%
%
\section*{Bibliography}
\bibliographystyle{abbrv} 
\bibliography{references}
\end{document}